\documentclass[onecolumn,nofootinbib,10pt,floatfix]{revtex4-2}
\usepackage[T1]{fontenc}
\usepackage{amsmath,amssymb,graphicx,microtype}
\usepackage{xr-hyper}
\usepackage[colorlinks=true,allcolors=black]{hyperref}
\hypersetup{pdftitle={Information geometry of DGLAP evolution},pdfauthor={Sergey Bondarenko and Alex Prygarin}}
\makeatletter
\@ifundefined{r@app:hera}{\newlabel{app:hera}{{S1}{1}{}{}{}}}{}
\@ifundefined{r@app:pole}{\newlabel{app:pole}{{S7}{10}{}{}{}}}{}
\@ifundefined{r@app:resummed}{\newlabel{app:resummed}{{S6}{8}{}{}{}}}{}
\@ifundefined{r@eq:memorysplit}{\newlabel{eq:memorysplit}{{S27}{11}{}{}{}}}{}
\@ifundefined{r@fig:hera}{\newlabel{fig:hera}{{S1}{6}{}{}{}}}{}
\@ifundefined{r@sm:approach}{\newlabel{sm:approach}{{S8}{11}{}{}{}}}{}
\@ifundefined{r@sm:fitmethods}{\newlabel{sm:fitmethods}{{S5}{6}{}{}{}}}{}
\@ifundefined{r@sm:heraresults}{\newlabel{sm:heraresults}{{S4}{5}{}{}{}}}{}
\@ifundefined{r@sm:scan}{\newlabel{sm:scan}{{S9}{13}{}{}{}}}{}
\makeatother
\renewcommand{\d}{\partial}
\newlength{\assumpnamewd}
\newlength{\assumpcontentwd}
\newlength{\assumpusewd}
\newlength{\titlepagetrim}
\newcommand{\NumMu}{1.4945}  
\newcommand{\NumD}{3.5893}  
\newcommand{\NumLambdaThree}{10.238}  
\newcommand{\NumDoverMu}{2.4016}  
\newcommand{\NumMuTau}{5.7290}  
\newcommand{\NumDriftLo}{3.3865}  
\newcommand{\NumDriftHi}{7.9252}  
\newcommand{\NumDriftLoZeta}{0.8834}  
\newcommand{\NumDriftHiZeta}{2.0674}  
\newcommand{\NumDriftHiNfThree}{7.6196}  
\newcommand{\NumDriftHiNfFour}{7.7724}  
\newcommand{\NumDriftHiNfSix}{8.0779}  
\newcommand{\NumTauConst}{1.5445}  
\newcommand{\NumAkappa}{3.024}  
\newcommand{\NumAL}{1.578}  
\newcommand{\NumMuNfThree}{1.3546}  
\newcommand{\NumDNfThree}{3.1501}  
\newcommand{\NumLambdaThreeNfThree}{7.658}  
\newcommand{\NumAkappaNfThree}{3.434}  
\newcommand{\NumALNfThree}{1.526}  
\newcommand{\NumTauConstNfThree}{1.6190}  
\newcommand{\NumMuNfFour}{1.4143}  
\newcommand{\NumDNfFour}{3.3623}  
\newcommand{\NumLambdaThreeNfFour}{8.962}  
\newcommand{\NumAkappaNfFour}{3.214}  
\newcommand{\NumALNfFour}{1.543}  
\newcommand{\NumTauConstNfFour}{1.5744}  
\newcommand{\NumMuNfSix}{1.5984}  
\newcommand{\NumDNfSix}{3.8520}  
\newcommand{\NumLambdaThreeNfSix}{11.544}  
\newcommand{\NumAkappaNfSix}{2.850}  
\newcommand{\NumALNfSix}{1.629}  
\newcommand{\NumTauConstNfSix}{1.5236}  
\newcommand{\NumCorrSqMin}{0.948}  
\newcommand{\NumCorrSqMax}{0.995}  
\newcommand{\NumCorrSqDenseMin}{0.677}  
\newcommand{\NumGaussCorrSqMin}{0.373}  
\newcommand{\NumGaussCorrSqMax}{0.762}  
\newcommand{\NumCorrSqHardqSteepg}{0.907}  
\newcommand{\NumCorrSqSoftqHardg}{0.721}  
\newcommand{\NumGaussSoftqHardg}{0.764}  
\newcommand{\NumCorrSqCTLOMin}{0.977}  
\newcommand{\NumCorrSqCTLOMax}{0.988}  
\newcommand{\NumAsymOverKappaTwo}{2.06}  
\newcommand{\NumAsymOverKappaTenTeV}{1.08}  
\newcommand{\NumZetaTenTeV}{1.71}  
\newcommand{\NumKappaWTenTeVA}{2.0838}  
\newcommand{\NumBTenTeVA}{2.1244}  
\newcommand{\NumMbStepLO}{1.26}  
\newcommand{\NumWinBelowJTwoHundred}{0.345}  
\newcommand{\NumWinBelowJTwoTenTeV}{1.343}  
\newcommand{\NumWinAboveJTwoTwo}{0.6}  
\newcommand{\NumWinAboveJTwoFive}{0.3}  
\newcommand{\NumWinAboveJTwoTen}{0.1}  
\newcommand{\NumKappaCTTwo}{4.54}  
\newcommand{\NumKappaCTUncTwo}{0.10}  
\newcommand{\NumKappaCTTen}{3.34}  
\newcommand{\NumKappaCTUncTen}{0.04}  
\newcommand{\NumKappaCTHundred}{2.79}  
\newcommand{\NumKappaCTUncHundred}{0.02}  
\newcommand{\NumKappaCTTenTeV}{2.34}  
\newcommand{\NumKappaCTUncTenTeV}{0.01}  
\newcommand{\NumKappaNNPDFTwo}{4.51}  
\newcommand{\NumKappaNNPDFTen}{3.33}  
\newcommand{\NumKappaNNPDFHundred}{2.78}  
\newcommand{\NumKappaNNPDFTenTeV}{2.34}  
\newcommand{\NumFallTwoTenTeV}{1.94}  
\newcommand{\NumFallShareBelowTen}{0.55}  
\newcommand{\NumMomShareTen}{15.4}  
\newcommand{\NumFisherShareTen}{53.6}  
\newcommand{\NumXminSensTenTeV}{1.72}  
\newcommand{\NumNNPDFSpreadTwo}{0.74}  
\newcommand{\NumNNPDFNusedTwo}{76}  
\newcommand{\NumNNPDFNtotTwo}{100}  
\newcommand{\NumNNPDFPosQsq}{5}  
\newcommand{\NumNNPDFCTMaxDiff}{0.5}  
\newcommand{\NumNNPDFCTRelTwo}{\ensuremath{-0.5}}  
\newcommand{\NumNNPDFCTRelTen}{\ensuremath{-0.3}}  
\newcommand{\NumNNPDFCTRelHundred}{\ensuremath{-0.3}}  
\newcommand{\NumNNPDFCTRelTenTeV}{\ensuremath{-0.2}}  
\newcommand{\NumCOnePrimeSigma}{\ensuremath{-10.5904}}  
\newcommand{\NumCOnePrimeG}{\ensuremath{-15.2841}}  
\newcommand{\NumCOnePrimeSigmaNfFour}{\ensuremath{-12.1300}}  
\newcommand{\NumCOnePrimeGNfFour}{\ensuremath{-18.5511}}  
\newcommand{\NumRNLOMax}{0.050}  
\newcommand{\NumRNLOScan}{0.061}  
\newcommand{\NumRNLOScanQ}{5.16}  
\newcommand{\NumRLOTwo}{13.47}  
\newcommand{\NumRLOFive}{7.96}  
\newcommand{\NumRLOTen}{6.59}  
\newcommand{\NumRLOHundred}{4.26}  
\newcommand{\NumFinitePartTwo}{10.9}  
\newcommand{\NumFinitePartHundred}{2.6}  
\newcommand{\NumResObsTwo}{\ensuremath{-5.91}}  
\newcommand{\NumResObsHundred}{\ensuremath{-2.31}}  
\newcommand{\NumHOneRate}{0.0481}  
\newcommand{\NumAOneScanN}{152}  
\newcommand{\NumAOneNNLOGridMax}{0.05}  
\newcommand{\NumLambdaQCD}{0.2}  
\newcommand{\NumAlphasMZModel}{0.134}  
\newcommand{\NumQuarkFracSteep}{0.55}  
\newcommand{\NumQuarkFracValence}{0.70}  
\newcommand{\NumInSteepQOnec}{0.30}  
\newcommand{\NumInSteepQOnea}{0.6}  
\newcommand{\NumInSteepQTwoc}{0.25}  
\newcommand{\NumInSteepQTwoa}{\ensuremath{-0.35}}  
\newcommand{\NumInSteepGc}{0.45}  
\newcommand{\NumInSteepGa}{\ensuremath{-0.4}}  
\newcommand{\NumInValenceQc}{0.70}  
\newcommand{\NumInValenceQa}{0.7}  
\newcommand{\NumInValenceGc}{0.30}  
\newcommand{\NumInValenceGa}{0.5}  
\newcommand{\NumInHqSgQc}{0.55}  
\newcommand{\NumInHqSgQa}{1}  
\newcommand{\NumInHqSgGc}{0.45}  
\newcommand{\NumInHqSgGa}{\ensuremath{-0.45}}  
\newcommand{\NumInSqHgQc}{0.30}  
\newcommand{\NumInSqHgQa}{\ensuremath{-0.3}}  
\newcommand{\NumInSqHgGc}{0.70}  
\newcommand{\NumInSqHgGa}{1}  
\newcommand{\NumThresholdScanLo}{1.5}  
\newcommand{\NumTalbotDigits}{50}  
\newcommand{\NumTalbotTerms}{48}  
\newcommand{\NumTalbotDigitsHigh}{100}  
\newcommand{\NumTalbotTermsHigh}{96}  
\newcommand{\NumBandFirstEnd}{4.58}  
\newcommand{\NumMbCT}{4.75}  
\newcommand{\NumCTHessianPairs}{29}  
\newcommand{\NumCTHessianCL}{90}  
\newcommand{\NumAlphasLow}{0.116}  
\newcommand{\NumAlphasHigh}{0.120}  
\newcommand{\NumThresholdScanN}{72}  
\newcommand{\NumInterpLowQ}{1.5}  
\newcommand{\NumHOneQSqMin}{3.5}  
\newcommand{\NumResWindowHi}{0.8}  
\newcommand{\NumCorrSqSoftqHardgQZero}{0.660}  
\newcommand{\NumRNLOTwo}{\ensuremath{-0.003}}  
\newcommand{\NumRNLOFive}{0.050}  
\newcommand{\NumRNLOTen}{\ensuremath{-0.008}}  
\newcommand{\NumRNLOHundred}{\ensuremath{-0.029}}  
\newcommand{\NumCTContXu}{0.98}  
\newcommand{\NumCTContPowShift}{0.1}  
\newcommand{\NumCTContSens}{\ensuremath{5.9\times10^{-5}}}  
\newcommand{\NumKjointRatioMin}{1.034}  
\newcommand{\NumKjointRatioMax}{1.082}  
\newcommand{\NumCorrSqSteepQZero}{0.940}  
\newcommand{\NumBandUncTwo}{2.17}  
\newcommand{\NumBandUncTen}{1.12}  
\newcommand{\NumBandUncHundred}{0.82}  
\newcommand{\NumBandUncThousand}{0.70}  
\newcommand{\NumBandUncTenTeV}{0.64}  
\newcommand{\NumZetaTen}{0.69}  
\newcommand{\NumZetaHundred}{1.16}  
\newcommand{\NumLambdaAlt}{0.088}  
\newcommand{\NumAlphasMZAlt}{0.118}  
\newcommand{\NumZetaAltTen}{0.53}  
\newcommand{\NumZetaAltHundred}{0.93}  
\newcommand{\NumZetaAltTenTeV}{1.43}  
\newcommand{\NumCorrSqSoftqHardgTen}{0.823}  
\newcommand{\NumCorrSqSoftqHardgHundred}{0.875}  
\newcommand{\NumCorrSqSoftqHardgTenTeV}{0.918}  
\newcommand{\NumCorrSqHardqSteepgMin}{0.907}  
\newcommand{\NumCorrSqHardqSteepgMax}{0.922}  
\newcommand{\NumCorrSqExtraMin}{0.721}  
\newcommand{\NumCorrSqExtraMax}{0.922}  
\newcommand{\NumCorrSqMinPct}{94.8}  
\newcommand{\NumCorrSqExtraMinPct}{72.1}  
\newcommand{\NumCorrSqExtraMaxPct}{92.2}  
\newcommand{\NumGeoLA}{1.261}  
\newcommand{\NumGeoDA}{1.226}  
\newcommand{\NumGeoBA}{1.253}  
\newcommand{\NumGeoExcessA}{2.9}  
\newcommand{\NumGeoExcessBA}{0.7}  
\newcommand{\NumGeoExcessMin}{1.7}  
\newcommand{\NumGeoExcessTwoMin}{1.4}  
\newcommand{\NumGeoExcessMax}{7.7}  
\newcommand{\NumGeoExcessTwoMax}{5.6}  
\newcommand{\NumGeoTiltExcessA}{6.6}  
\newcommand{\NumGeoTiltExcessMin}{4.0}  
\newcommand{\NumGeoTiltExcessMax}{11.5}  
\newcommand{\NumGeoTiltDistA}{0.28}  
\newcommand{\NumGeoTiltDistSteep}{0.34}  
\newcommand{\NumGeoCurvLOTwoA}{0.36}  
\newcommand{\NumGeoCurvTiltTwoA}{0.82}  
\newcommand{\NumGeoCurvLOTenTeVA}{0.47}  
\newcommand{\NumGeoCurvTiltTenTeVA}{0.60}  
\newcommand{\NumScanN}{200}  
\newcommand{\NumScanMinQZero}{0.695}  
\newcommand{\NumScanMaxQZero}{0.998}  
\newcommand{\NumScanMinTwo}{0.802}  
\newcommand{\NumScanMaxTwo}{0.999}  
\newcommand{\NumScanMinTenTeV}{0.941}  
\newcommand{\NumScanMaxTenTeV}{0.985}  
\newcommand{\NumScanCountNinetyTen}{198}  
\newcommand{\NumRegDefSmallTwoA}{65}  
\newcommand{\NumRegMomSmallTwoA}{3}  
\newcommand{\NumRegDefLargeTenTeVA}{45}  
\newcommand{\NumRegMomLargeTenTeVA}{6}  
\newcommand{\NumCorrSqValenceTwo}{0.982}  
\newcommand{\NumTransTauHalf}{0.00579}  
\newcommand{\NumTransQsqHalf}{2.18}  
\newcommand{\NumTransRatio}{50}  
\newcommand{\NumTransR}{35.9}  
\newcommand{\NumTransA}{5.28}  
\newcommand{\NumTransCoef}{3330}  
\newcommand{\NumLamRateRatioA}{0.44}  
\newcommand{\NumNNPDFSpreadTen}{0.32}  
\newcommand{\NumNNPDFNusedTen}{75}  
\newcommand{\NumNNPDFSpreadHundred}{0.22}  
\newcommand{\NumNNPDFNusedHundred}{75}  
\newcommand{\NumNNPDFSpreadTenTeV}{0.17}  
\newcommand{\NumNNPDFNusedTenTeV}{75}  
\newcommand{\NumRegEdgeSmall}{0.0025}  
\newcommand{\NumRegEdgeLarge}{0.22}  
\newcommand{\NumGeoBestExcessMax}{7.7}  
\newcommand{\NumNullSqrtXiMin}{0.959}  
\newcommand{\NumNullSqrtXiMax}{0.968}  
\newcommand{\NumNullXiSqMin}{0.86}  
\newcommand{\NumNullXiSqMax}{0.90}  
\newcommand{\NumNullLnXiMin}{0.83}  
\newcommand{\NumNullLnXiMax}{0.87}  
\newcommand{\NumNullLnOneMinusXMin}{0.29}  
\newcommand{\NumNullLnOneMinusXMax}{0.57}  
\newcommand{\NumSlopeSqrtXiHighMin}{0.991}  
\newcommand{\NumSlopeSqrtXiHighMax}{0.998}  
\newcommand{\NumSlopeXiHighMin}{0.962}  
\newcommand{\NumSlopeXiHighMax}{0.985}  
\newcommand{\NumSlopePStarHighMin}{0.49}  
\newcommand{\NumSlopePStarHighMax}{0.71}  
\newcommand{\NumSlopeSqrtBetterCount}{10}  
\newcommand{\NumSlopePScanLo}{0.05}  
\newcommand{\NumSlopePScanHi}{3}  

\begin{document}
\title{Information geometry of DGLAP evolution}
\author{Sergey Bondarenko}
\email{sergeyb@ariel.ac.il}
\affiliation{Department of Physics, Ariel University, Ariel 40700, Israel}
\author{Alex Prygarin}
\email{alexanderp@ariel.ac.il}
\affiliation{Department of Physics, Ariel University, Ariel 40700, Israel}

\begin{abstract}
Quark and gluon splittings shift proton momentum toward smaller fractions with increasing resolution.
Momentum conservation makes the summed momentum-weighted quark and gluon densities, when nonnegative, a probability distribution.
We apply information geometry, the geometry of probability distributions, to this probabilistic form of Dokshitzer-Gribov-Lipatov-Altarelli-Parisi evolution.
We evaluate, to our knowledge for the first time, the speed at which the shape of the proton's momentum distribution changes, measured by how well distributions at neighboring resolutions can be told apart statistically.
We show that this speed has a closed-form lower bound at leading order, with one-loop splitting functions and running coupling, five fixed quark flavors, nonnegative quark and gluon densities and sufficiently regular inputs.
Three numbers of the input (the distribution at the starting resolution) fix the bound at every scale without computing the evolved distribution, once the coupling is specified.
The squared bound is at least \NumCorrSqMinPct{} percent of the squared speed for three of the five model inputs at four scales from 2 GeV to 10 TeV.
The bound is reached when the distribution's relative rate of change is linear in the logarithm of the inverse momentum fraction.
The relative rate of change is close to this linear form in mean square weighted by momentum for these three model inputs at all four scales, and closer still to linear in the square root of the logarithm at the three higher scales.
These descriptions are compatible because the logarithm and its square root are strongly correlated when weighted by these momentum distributions.
\end{abstract}
\maketitle
\vspace*{-\titlepagetrim}

\section{Introduction}
\label{sec:intro}

Parton splittings redistribute the proton's momentum as a scattering process resolves shorter distances.
The Dokshitzer-Gribov-Lipatov-Altarelli-Parisi (DGLAP) equations~\cite{GribovLipatov1972,Lipatov1975,AltarelliParisi1977,Dokshitzer1977} describe this scale dependence.
Scaling violations of structure functions describe this change at each momentum fraction separately.
A single rate for the whole distribution would quantify the difference between neighboring resolutions, weighted by momentum.
Momentum conservation normalizes the sum of the momentum-weighted quark and gluon densities to one.
This sum is a probability distribution in momentum fraction when it is nonnegative at every momentum fraction.

H1~\cite{H1rise2001} described the rise of the proton structure function $F_2$ at small momentum fractions $x$ by a single power whose exponent increased with resolution.
Prytz~\cite{Prytz1993} related the small-$x$ scale derivative of $F_2$ to the gluon momentum density at leading order (LO), with one-loop splitting functions.
H1~\cite{H1lowx2001} measured this derivative's rise toward small $x$ at fixed $Q^2$.
ZEUS~\cite{ZEUS1999} found slopes rising as $x$ fell, with a turnover in bins reaching low $Q^2$.
Those slopes came from fits linear in $\ln Q^2$ at fixed $x$, an approximation H1~\cite{H1lowx2001} found inconsistent with its data.
Ball, Nocera and Rojo~\cite{BallNoceraRojo2016} also followed effective exponents of fitted densities with scale.
These local descriptions leave open how much of the whole distribution's shape change one effective power describes.

Benito-Calvi\~no, Garc\'{\i}a-Olivares and Llanes-Estrada~\cite{BenitoCalvino2023,BenitoCalvino2022} used momentum weighting to normalize parton distributions and fragmentation functions.
Both works studied Shannon entropy, which measures spread, and relative entropy, which measures how well distributions can be told apart.
The relative-entropy comparison in Ref.~\cite{BenitoCalvino2023} found that a quark-model relation between pion parton distributions and fragmentation functions is not well satisfied.

Jadach and Skrzypek~\cite{JadachSkrzypek2004} described LO DGLAP evolution as a Markov process, as did Teryaev~\cite{Teryaev1998}, and solved it by Monte Carlo sampling~\cite{JadachSkrzypek2004}. The next step of a Markov process depends only on its present state.
At LO, the logarithmic momentum loss and the species label form a Markov additive process (Appendix~\ref{app:markov}), as in the multitype fragmentation described by Bertoin~\cite{Bertoin2008}, where a fragment can change type as the logarithm of its mass decreases.
Stochastic descriptions also exist for dipole or Pomeron numbers in the Balitsky-Fadin-Kuraev-Lipatov (BFKL) Pomeron calculus in work by Kozlov, Levin and Prygarin~\cite{KozlovLevinPrygarin2006} and by Levin and Prygarin~\cite{LevinPrygarin2007} and, as a L\'evy process, for normalized forward BFKL evolution at LO and fixed coupling in the work of Prygarin, Madjuogang Sandeu and Shekh Yusuf~\cite{PrygarinSandeuYusuf2026}.
Golec-Biernat~\cite{GolecBiernat2025} derived how the momentum density's relative entropy to the uniform distribution changes with $\ln Q^2$.
That work also evaluated relative entropy between evolved and input densities at finite scale separation~\cite{GolecBiernat2025}.
It did not take the small-separation limit or introduce Fisher information, speed or length.

Kharzeev and Levin~\cite{KharzeevLevin2017} proposed an entanglement entropy for parton numbers at small $x$, arising from correlations with unobserved partons, and related it to final-state hadron entropy.
H1~\cite{H1entropy2021} measured charged-particle multiplicity distributions and their entropies in deep-inelastic scattering at HERA.
Hentschinski and Kutak~\cite{HentschinskiKutak2022} included quarks and reported agreement with the H1 entropy for a gluon distribution retaining transverse momentum with next-to-leading-order (NLO) BFKL evolution.
The corrected comparison in the erratum of Hentschinski and Kutak~\cite{HentschinskiKutak2023E} used a fixed-order fit with three-loop splitting functions and a small-$x$ resummed fit by Ball et~al.~\cite{BallResum2018}.
The fixed-order entropy lay below the H1 result at small $x$, while the resummed entropy followed its $x$ dependence~\cite{HentschinskiKutak2023E}.
Hentschinski, Kharzeev, Kutak and Tu~\cite{HKKT2025} studied entanglement-entropy evolution in quantum chromodynamics (QCD) and extended it to rapidity windows.

Peschanski~\cite{Peschanski2013} obtained the quadratic limit of relative entropy between gluon distributions at different rapidities in a Gaussian-tail model.
Ramos, Moriggi and Machado~\cite{Ramos2025} identified the quadratic coefficient for a generic density family as the Fisher-Rao metric, but did not evaluate it in their rapidity application.
Ramos and Machado~\cite{RamosMachado2022} also studied dynamical entropy in hadronic collisions.
Other weightings occur for transverse phase-space entropy studied by Hagiwara, Hatta, Xiao and Yuan~\cite{HHXY2018}, parton-distribution entropy studied by Wang~\cite{Wang2024}, nuclear versus free-nucleon relative entropy studied by Hu et~al.~\cite{HuXiongXu2026}, and small-$x$ deep-inelastic entropy evolved by DGLAP in the work of Boroun and Ha~\cite{BorounHa2026}.
These studies use different densities or weightings, or evolve in rapidity instead of $\ln Q^2$.

Information metrics also describe renormalization-group flows.
Dolan~\cite{Dolan1998} related a metric on the couplings of the large-$N$ $O(N)$ model to infinitesimal relative entropy and defined curve length.
Brody and Ritz~\cite{BrodyRitz1998} evaluated the Fisher-Rao speed of real-space renormalization-group flow for the one-dimensional Ising model in a magnetic field.
Balasubramanian, Heckman and Maloney~\cite{BHM2015} related the second derivative of relative entropy at coincident logarithmic scales to the squared rate of change of couplings in a two-dimensional conformal field theory.
B\'eny and Osborne~\cite{BenyOsborne2015} also considered renormalization-group speed.
These paths are traced by couplings of a field theory or spin model, not by a parton distribution.

Entropy describes spread at one scale.
Fisher information determines the quadratic growth of relative entropy between neighboring scales for smooth positive densities with the regularity stated at Eq.~\eqref{eq:kl}, as in Ito and Dechant~\cite{ItoDechant2020} and in Ramos, Moriggi and Machado~\cite{Ramos2025}.
The square root of the Fisher information is the speed.
In this paper we apply information geometry, the geometry of probability distributions with the Fisher-Rao metric~\cite{Fisher1922,Rao1945}, to the probabilistic form of LO DGLAP evolution.
To our knowledge, we evaluate for the first time the Fisher information and speed of the momentum density $P=x\Sigma+xg$ along $\ln Q^2$, where $\Sigma$ is the quark singlet and $g$ the gluon density.
Kou and Chen~\cite{KouChen2026} independently computed the Fisher-Rao speed, path length, its excess over the distance between the end densities, and geodesic curvature for the LO non-singlet evolution of a normalized valence-quark number density.
Their work, posted on arXiv on the same day as this paper's e-print, found that the path is not a geodesic and, over the range of scales they evolved, stays close to a small circle, which has constant nonzero geodesic curvature.
We treat the momentum density of coupled quark and gluon evolution and derive a lower bound on its speed.

The speed divided by $\alpha_s/(2\pi)$ is $\kappa$, the speed per unit of the evolution time $\tau$ of Eq.~\eqref{eq:clocks}.
Let $B$ be the absolute rate of change of the mean of $\xi=\ln(1/x)$ per unit $\tau$, divided by the standard deviation of $\xi$.
Ito and Dechant~\cite{ItoDechant2020} and Nicholson, Garc\'{\i}a-Pintos, del Campo and Green~\cite{Nicholson2020} used the Cauchy-Schwarz inequality in stochastic thermodynamics. This inequality gives $\kappa\ge B$ if the mean-bound assumptions of Table~\ref{tab:assumptions} hold.
These include exchanging the scale derivative with the normalization and first logarithmic moment integrals, checked on the evaluated inputs but not proved for every admissible LO input.
The speed is not an observable, since it depends on the factorization scheme, the perturbative order, the normalization interval and, near thresholds, the derivative prescription (Sec.~\ref{sec:density}~B).

The structure of this paper is as follows.
Sections~\ref{sec:density} through~\ref{sec:markov} develop the density, moments, bounds and path geometry.
Section~\ref{sec:fitted} evaluates fitted densities.
The appendices give derivations and evaluation methods.
The Supplemental Material~\cite{SupplementalMaterial} gives the structure-function speed, fitted-density comparisons, input scan and further geometry.

The main result of this paper is the closed-form lower bound $B$ of Eq.~\eqref{eq:Bzeta} on the speed per unit evolution time in LO DGLAP evolution with one-loop splitting functions, one-loop running coupling, five fixed quark flavors, separately nonnegative quark and gluon inputs, the LO moment and mean-bound assumptions of Table~\ref{tab:assumptions}, and positive variance of $\xi$ at every evaluated scale.
The bound needs only the input quark momentum fraction, centered quark moment and variance of $\xi$, together with the coupling, at every value of the evolution time without solving for the evolved density.
The centered quark moment is the quark part of the centered first moment of $\xi$.
If both momentum fractions are nonzero, it is their product times the difference between the quark and gluon means of $\xi$.
LO evolution preserves the separate nonnegativity of the input densities, so $B$ bounds $\kappa$ at every scale of the LO path if the other mean-bound assumptions hold.
The squared ratio $(B/\kappa)^2$ ranges from $\NumCorrSqMin$ to $\NumCorrSqMax$ at 2, 10, 100 and $10^4$~GeV for three of the five model inputs of Appendix~\ref{app:methods}, proton-like, steep at small $x$ and valence-like with no small-$x$ rise.
A separate LO trajectory from the CT18 global fit of the CTEQ-TEA group~\cite{CT18} at next-to-next-to-leading order (NNLO), with the coupling of that fit and a bottom-quark threshold, gives $\NumCorrSqCTLOMin$ to $\NumCorrSqCTLOMax$ at 2, 5, 10, 100 and $10^4$~GeV, with $B$ evaluated from the evolved density because the threshold prevents use of the fixed five-flavor closed form.
The geometry gives each evolution path a length, and the five model paths from the input scale to $10^4$~GeV are $\NumGeoExcessMin$ to $\NumGeoExcessMax$ percent longer than the shortest paths between their end densities (Sec.~\ref{sec:bound}~C).
The speed of $F_2$ normalized on a fixed window equals its mean bound for any exponent if $F_2$ has the single-power form used by H1~\cite{H1rise2001}.
This equality is a property of the form, not of the H1 data.
The LO logarithmic rate of change of the evolved momentum density is close in momentum-weighted mean square to a slope affine in $\ln(1/x)$ for these three inputs at the four scales.
This rate of change is closer still to an affine function of $\sqrt{\ln(1/x)}$ for these inputs at 10, 100 and $10^4$~GeV and can be close to both forms because $\sqrt{\ln(1/x)}$ and $\ln(1/x)$ are strongly correlated when weighted by these momentum densities (Sec.~\ref{sec:bound}).
The bound computed from three numbers of the input, with the coupling specified, is within a few percent of the speed obtained from the evolved density for these three inputs at the four scales.

\section{The momentum density and its speed}
\label{sec:density}

Momentum weighting turns the distribution into a probability law. The quark singlet and gluon parts are $P_q=x\Sigma$ and $P_g=xg$, with~\cite{BenitoCalvino2023,GolecBiernat2025}
\begin{equation}
\begin{gathered}
P(x,Q^2)=\sum_i x f_i(x,Q^2)=x\Sigma+xg,\\
\int_0^1 P(x,Q^2)\,dx=1 .
\end{gathered}
\label{eq:Pdef}
\end{equation}
The measure is $dx$. In $\xi=\ln(1/x)$, the density $p_j(\xi)=xP_j(x)$ includes the Jacobian needed to preserve the probability law. Normalization with a different measure would define a different law.

The evolution time separates the splitting dynamics from the running of the coupling. We use
\begin{equation}
\begin{gathered}
t=\ln Q^2,\qquad
\tau=\int_{t_0}^{t}\frac{\alpha_s(t')}{2\pi}\,dt',\\
\zeta=\ln\frac{T}{T_0},\qquad
T=\ln\frac{Q^2}{\Lambda_{\rm QCD}^2},\qquad
\alpha_s=\frac{1}{b_0T},\\
b_0=\frac{\beta_0}{4\pi},\qquad
\beta_0=11-\frac{2n_f}{3},\qquad
\frac{d\zeta}{d\tau}=2\pi b_0=\frac{\beta_0}{2}.
\end{gathered}
\label{eq:clocks}
\end{equation}
The definition of $\tau$ requires only a positive coupling, and the LO statements in $\tau$ hold for any positive coupling. The displayed $\alpha_s$ and constant conversion to $\zeta$ require one-loop running at fixed flavor number. Both times vanish at $Q_0$, with $t_0=\ln Q_0^2$ and $T_0=\ln(Q_0^2/\Lambda_{\rm QCD}^2)$. For the model inputs, $Q_0^2=2$~GeV$^2$ and $\Lambda_{\rm QCD}=\NumLambdaQCD$~GeV give $\alpha_s(M_Z)=\NumAlphasMZModel$ with one-loop running and five flavors fixed up to $M_Z$. The $n_f=5$ model curves serve as illustrations of the LO closed forms, and the LO trajectory started from the CT18 global fit of the CTEQ-TEA group at NNLO~\cite{CT18} is the realistic case. Fixed $n_f=5$ is a convention that lets one closed form describe each model curve over its full range, not the physical flavor content at $Q_0$ below the bottom threshold. Appendix~\ref{app:matrix} gives the flavor dependence. Thresholds are treated by the LO matching of Eqs.~\eqref{eq:thresholdjump} and~\eqref{eq:thresholddrift} and on the LO trajectory from CT18~\cite{CT18} in Appendix~\ref{app:methods}. That appendix also gives the effect of changing $\Lambda_{\rm QCD}$. A speed changes with the time coordinate, while the length of the same path does not.

\subsection{Assumptions}

The moment equations determine the drift and width. The information bounds also require a probability density and control of its scale derivative. Table~\ref{tab:assumptions} collects the assumptions and the results that use them.

\begin{table*}[t]
\caption{Assumptions for the moment equations and information bounds. Each result uses the named rows indicated in the text.}
\label{tab:assumptions}
\begin{ruledtabular}
\begin{tabular}{p{\assumpnamewd}p{\assumpcontentwd}p{\assumpusewd}}
Name & Content & Used for\\
\hline
{\raggedright LO moment assumptions\par} & {\raggedright Fixed flavor number, one-loop singlet kernel, full interval $0<x<1$, unit momentum sum, finite absolute logarithmic moments of the orders used, and differentiation under their integrals. The drift uses the first moment and the width the second. The drift identity needs no positivity.\par} & {\raggedright Drift, mean, variance, and $B$ where the variance is positive, Eqs.~\eqref{eq:drift}, \eqref{eq:meanxi}, \eqref{eq:varxi}, and \eqref{eq:Bzeta}, and their large-time forms. The drift interval of Eq.~\eqref{eq:driftinterval} holds at every scale of a forward LO path if and only if the input quark momentum fraction lies in $[0,1]$.\par}\\ \noalign{\vskip 4pt}
{\raggedright LO positivity assumption\par} & {\raggedright Separately nonnegative $P_q$ and $P_g$ at the input. LO evolution preserves them (Sec.~\ref{sec:markov}). The joint gap and the right inequality of Eq.~\eqref{eq:window} also require finite joint information.\par} & {\raggedright Joint gap, Eq.~\eqref{eq:jointgap}, right inequality of Eq.~\eqref{eq:window}, and $\kappa\ge B$, Eq.~\eqref{eq:identity}, with $B$ of Eq.~\eqref{eq:Bzeta} on LO paths.\par}\\ \noalign{\vskip 4pt}
{\raggedright Fisher assumptions\par} & {\raggedright A density nonnegative throughout a fixed normalization interval, differentiable in scale almost everywhere, with finite information and differentiation under the normalization integral. Zeros are allowed when the information integrals are defined by their limits. Beyond LO, positivity must be checked for the density and interval used.\par} & {\raggedright Information, speed, and length, Eqs.~\eqref{eq:vLdef}, \eqref{eq:Ldef}, and \eqref{eq:var}, and window split, Eq.~\eqref{eq:windowsplit}, with $\d_\tau Z_W=\int_W\d_\tau P\,dx$.\par}\\ \noalign{\vskip 4pt}
{\raggedright Derivative exchange ($H_{\rm ex}$)\par} & {\raggedright Differentiation under the normalization and first logarithmic moment integrals. Checked on the evaluated inputs but not proved for every admissible LO input.\par} & {\raggedright Mean bound, Eq.~\eqref{eq:identity}, and threshold matching, Eqs.~\eqref{eq:thresholdjump} and \eqref{eq:thresholddrift}.\par}\\ \noalign{\vskip 4pt}
{\raggedright Mean-bound assumptions\par} & {\raggedright The Fisher assumptions, $H_{\rm ex}$, and finite nonzero logarithmic variance. The correlation form also requires positive information.\par} & {\raggedright Eqs.~\eqref{eq:covdrift} and \eqref{eq:identity}.\par}\\ \noalign{\vskip 4pt}
{\raggedright Quadratic-moment assumptions\par} & {\raggedright The mean-bound assumptions, finite fourth logarithmic moment, differentiation under the second-moment integral, positive information, nonzero drift, and invertible covariance of $\xi$ and $\xi^2$.\par} & {\raggedright Eqs.~\eqref{eq:Jtwo} and \eqref{eq:Rlow} of Appendix~\ref{app:bound}.\par}\\ \noalign{\vskip 4pt}
{\raggedright Markov-contraction assumptions\par} & {\raggedright LO positivity, finite joint information, and differentiation under the transition integral.\par} & {\raggedright Joint-speed contraction, Eq.~\eqref{eq:jointmono}.\par}\\ \noalign{\vskip 4pt}
{\raggedright LO matching assumptions\par} & {\raggedright Continuous $P$ and $\alpha_s$, zero initial heavy density, finite total first logarithmic moment, $H_{\rm ex}$, and finite species-resolved information at the input.\par} & {\raggedright Flavor thresholds, Eqs.~\eqref{eq:thresholdjump} and \eqref{eq:thresholddrift}.\par}\\ \noalign{\vskip 4pt}
{\raggedright NLO moment assumptions\par} & {\raggedright A fixed-flavor segment in $\overline{\rm MS}$, both factorization and renormalization scales equal to $Q$, the full interval, finite first absolute logarithmic moments of each species, and differentiation under their integrals.\par} & {\raggedright NLO drift, Eq.~\eqref{eq:nlodrift}, and Appendix~\ref{app:nlo}.\par}\\
\end{tabular}
\end{ruledtabular}
\end{table*}

A scale-independent integrable majorant for the absolute derivative, with the corresponding moment weight, suffices locally for each derivative exchange. For a small-$x$ power, the bound $Cx^{-\bar\lambda}[1+\ln(1/x)]$, with $\bar\lambda<1$ bounding the exponent near $t$, suffices for normalization differentiation. Appendix~\ref{app:bound} treats the moment weights and endpoint terms. The relative-entropy expansion and second normalization derivative below require their indicated extra regularity.

\subsection{Information and shape change}

Relative entropy measures the difference between neighboring densities. If the family is smooth and positive on the fixed interval and has enough regularity for the remainder, relative entropy has the small-scale-separation expansion~\cite{ItoDechant2020,Ramos2025}
\begin{equation}
D_{\rm KL}\!\left[P(t+h)\Vert P(t)\right]
=\frac{h^2}{2}\int_0^1 dx\,\frac{\bigl(\d_tP\bigr)^2}{P}+O(h^3).
\label{eq:kl}
\end{equation}
The coefficient defines the Fisher information and speed along the trajectory,
\begin{equation}
I_F(t)=\int_0^1 dx\,\frac{\bigl(\d_tP\bigr)^2}{P},
\qquad
v(t)=\sqrt{I_F(t)} .
\label{eq:vLdef}
\end{equation}
For densities with zeros, $I_F$ is defined by the integral of Eq.~\eqref{eq:vLdef} over the set where $P>0$. The expansion of Eq.~\eqref{eq:kl} can then fail, and for the density of Eq.~\eqref{eq:vanishingend} with growing $x_{\max}$ the forward relative entropy is infinite.

We normalize the Fisher-Rao metric as $ds^2=I_F\,dt^2$. This omits the factor $1/4$ used when distance is measured as an angle between the amplitudes $\sqrt P$. The length $L$ is the information length of Nicholson and Kim~\cite{NicholsonKim2015} and of Kim~\cite{Kim2018}, the statistical length of Wootters~\cite{Wootters1981} up to that normalization, and the thermodynamic length of Crooks~\cite{Crooks2007}. The accumulated length is~\cite{ItoDechant2020}
\begin{equation}
L(t_1,t_2)=\int_{t_1}^{t_2}v(t)\,dt .
\label{eq:Ldef}
\end{equation}
A scale-independent invertible change from $x$ to $\xi$ leaves the information unchanged. The larger coordinate range at small $x$ does not increase its probability weight. Reparametrizing scale along the same path leaves $L$ unchanged.

The coupling-divided speed is
\begin{equation}
\begin{gathered}
\d_tP=\frac{\alpha_s}{2\pi}\,K_{\rm tot},\qquad
I_\tau=\int_0^1 dx\,\frac{K_{\rm tot}^2}{P},\\
\kappa=\frac{2\pi v}{\alpha_s}=\sqrt{I_\tau}=2\pi b_0\sqrt{I_\zeta}.
\end{gathered}
\label{eq:kappadefn}
\end{equation}
The last equality uses one-loop running. The speed $\kappa$ is not an observable. It depends on factorization scheme, perturbative order, normalization interval, and, for fitted grids near thresholds, the scale-derivative prescription. The quantity $K_{\rm tot}$ is the full scale derivative with the displayed coupling factor removed. At LO,
\begin{equation}
K_{\rm tot}^{\rm LO}(x,t)
=(1,1)\left[z\mathbf P^{(0)}\otimes\begin{pmatrix}x\Sigma\\ xg\end{pmatrix}\right](x).
\label{eq:ktot}
\end{equation}
Here $\mathbf P^{(0)}$ is the singlet splitting matrix and $\otimes$ denotes Mellin convolution. The quantity $K_{\rm tot}$ for a higher-order fitted density is obtained from that density's own scale derivative. Applying the LO kernel to that density is a separate operation.

The speed measures how much the scaling-violation slope varies across momentum fractions. Define
\begin{equation}
s(x,t)=\frac{\d\ln P(x,t)}{\d\ln Q^2}.
\label{eq:sdef}
\end{equation}
If the Fisher assumptions hold, the momentum sum rule gives~\cite{GolecBiernat2025}
\begin{equation}
\langle s\rangle_P=\int_0^1 dx\,P(x,t)\,s(x,t)=0 .
\label{eq:smean}
\end{equation}
The information is therefore~\cite{ItoDechant2020}
\begin{equation}
I_F(t)=\operatorname{Var}_P[s]=\bigl\langle s^2\bigr\rangle_P .
\label{eq:var}
\end{equation}
Normalization removes a common change in the magnitude of a density. Its speed therefore measures shape change, weighted by the momentum probability. A large local slope contributes strongly only where the density gives it weight. Neither a slope at one selected $x$ nor the mean slope of an unnormalized density determines this variance.

If a second derivative can also be taken under the normalization integral,
\begin{equation}
\bigl\langle s^2\bigr\rangle_P+\bigl\langle\d_ts\bigr\rangle_P=0 .
\label{eq:secondnorm}
\end{equation}

A changing small-$x$ exponent describes the shape evolution. Define the local exponent by
\begin{equation}
\xi=\ln\frac1x,
\qquad
\lambda(x,t)=\frac{\d\ln P(x,t)}{\d\xi}.
\label{eq:lambdadef}
\end{equation}
A single power over the whole fixed window gives
\begin{equation}
P_W(x,t)=\frac{x^{-\lambda(t)}}{\int_W dx\,x^{-\lambda(t)}},\qquad x\in W .
\label{eq:powerwindow}
\end{equation}
Normalization removes the mean logarithmic slope,
\begin{equation}
\d_t\ln P_W(x,t)=\dot\lambda\left[\ln\frac1x-\left\langle\ln\frac1x\right\rangle_{P_W}\right].
\label{eq:windowslope}
\end{equation}
The dot denotes $d/dt$, and the speed is
\begin{equation}
v_W=\left|\frac{d\lambda}{d\ln Q^2}\right|\operatorname{sd}_{P_W}\!\left[\ln\frac1x\right].
\label{eq:windowspeed}
\end{equation}
Here $\operatorname{sd}_{P_W}$ is the standard deviation under $P_W$. The exponent rate is multiplied by the width in logarithmic momentum. A local power does not establish this result for a general density because Eq.~\eqref{eq:powerwindow} assumes a common exponent throughout the window.

The input-pole approximation of Ellis, Kunszt, and Levin~\cite{EllisKunsztLevin1994}, which relates the slope at small $x$ to the larger singlet anomalous dimension, is compared with the LO-kernel speed in Sec.~\ref{sm:fitmethods} of the Supplemental Material. The drift and width below use the full LO matrix and need no small-$x$ approximation.

\section{Drift and width at \texorpdfstring{$N=2$}{N=2}}
\label{sec:drift}

Each splitting moves momentum toward smaller $x$, at different mean rates for quarks and gluons. Momentum conservation makes the total drift a combination of those rates with the species momentum fractions as coefficients, an average when the quark momentum fraction lies in $[0,1]$. The mean drift needs no detailed shape, although the difference of the species mean logarithmic momenta enters the width.

Mellin derivatives give the drift and width without reconstructing the density in $x$. Mellin derivatives were used for logarithmic cumulants in time-like, number-weighted jet spectra by Fong and Webber~\cite{FongWebber1991}, Lupia and Ochs~\cite{LupiaOchs1996}, Khoze and Ochs~\cite{KhozeOchs1997}, and P\'erez-Ramos and d'Enterria~\cite{PerezRamosdEnterria2014} and in the space-like non-singlet application of Artru et~al.~\cite{Artru2009}. General multistate formulas are given at finite time by Landi, Kewming, Mitchison and Potts in Eqs.~(86)--(87) of Ref.~\cite{Landi2024} and for zero-frequency cumulants by Flindt, Novotn\'y, Braggio and Jauho~\cite{Flindt2010}. The new results are the space-like, momentum-weighted drift interval of Eq.~\eqref{eq:driftinterval} and the finite-scale variance terms $V_\infty$, $V_E$, and $V_{\zeta E}$ of Eq.~\eqref{eq:varxi}, relative to time-like number-weighted cumulants~\cite{FongWebber1991,LupiaOchs1996,KhozeOchs1997,PerezRamosdEnterria2014} and the double-asymptotic scaling of De R\'ujula et~al.~\cite{DeRujula1974} and of Ball and Forte~\cite{BallForte1994}.

The Mellin equations express this reduction as
\begin{equation}
\begin{gathered}
\tilde P_j(N,\tau)=\int_0^1 dx\,x^{N-2}P_j(x,\tau),\\
\d_\tau\tilde{\mathbf P}(N,\tau)=\Gamma(N)\,\tilde{\mathbf P}(N,\tau),\\
\bigl\langle e^{u\xi}\bigr\rangle_P=(1,1)\,\tilde{\mathbf P}(2-u,\tau).
\end{gathered}
\label{eq:mellinevol}
\end{equation}
The LO space-like singlet matrix $\Gamma$ is given in Appendix~\ref{app:matrix}. Differentiation at $N=2$ gives logarithmic moments. The generating function is used where it exists, but the finite-order moment equations need only the corresponding finite moments, without an exponential moment of the input.

The larger eigenvalue determines the limiting cumulant rates. In one-loop $\zeta$ units, define
\begin{equation}
\Lambda(u)=\frac{\gamma_+^{(0)}(2-u)}{2\pi b_0},
\qquad
\Lambda(0)=0 .
\label{eq:Lambda}
\end{equation}
Its first three derivatives are
\begin{equation}
\mu=\Lambda'(0),
\qquad
D=\Lambda''(0),
\qquad
\Lambda_3=\Lambda'''(0).
\label{eq:muD}
\end{equation}
Momentum weighting fixes the point of evaluation and the coupled space-like system here.

If the LO moment assumptions hold, differentiation at $N=2$ gives
\begin{equation}
\begin{gathered}
\frac{d\langle\xi\rangle}{d\tau}=-c_q'(2)\,\langle x\rangle_q-c_g'(2)\,\langle x\rangle_g,\\
\langle x\rangle_j=\int_0^1 dx\,P_j,\qquad
c_j(N)=\sum_i\Gamma_{ij}(N).
\end{gathered}
\label{eq:drift}
\end{equation}
The column sums vanish at this point, so the species logarithmic moments cancel. Only the momentum fractions multiply the rates
\begin{equation}
c_q'(2)=1-\frac{4\pi^2}{9},
\qquad
c_g'(2)=\frac{65}{24}-\pi^2-\frac{11n_f}{72}.
\label{eq:cprime}
\end{equation}
Each negative column derivative is the mean logarithmic loss associated with momentum in its incoming species.

At a given scale, the drift is a convex combination of $-c_q'(2)$ and $-c_g'(2)$ if and only if the quark momentum fraction lies in $[0,1]$. In that case, for $n_f=5$,
\begin{equation}
\frac{4\pi^2}{9}-1\;\le\;\frac{d\langle\xi\rangle}{d\tau}\;\le\;\pi^2-\frac{35}{18}.
\label{eq:driftinterval}
\end{equation}
The edges are $\NumDriftLo$ and $\NumDriftHi$ per unit $\tau$. Pure-quark or pure-gluon input can reach an edge initially, after which species mixing puts the drift between them. Each active flavor raises the upper edge by $11/72$. The lower edge is independent of $n_f$ because $c_q'(2)$ contains no $n_f$ (Appendix~\ref{app:matrix}). If the LO moment assumptions hold, the interval holds along a forward LO path if and only if the input quark momentum fraction lies in $[0,1]$, independently of the small-$x$ exponent or large-$x$ profile of the input. By Eq.~\eqref{eq:xqrelax} below, the quark momentum fraction stays between its input value and $15/31$. Separate nonnegativity of the densities is sufficient but not necessary.

\begin{figure*}[t]
\centering
\includegraphics[width=0.62\textwidth]{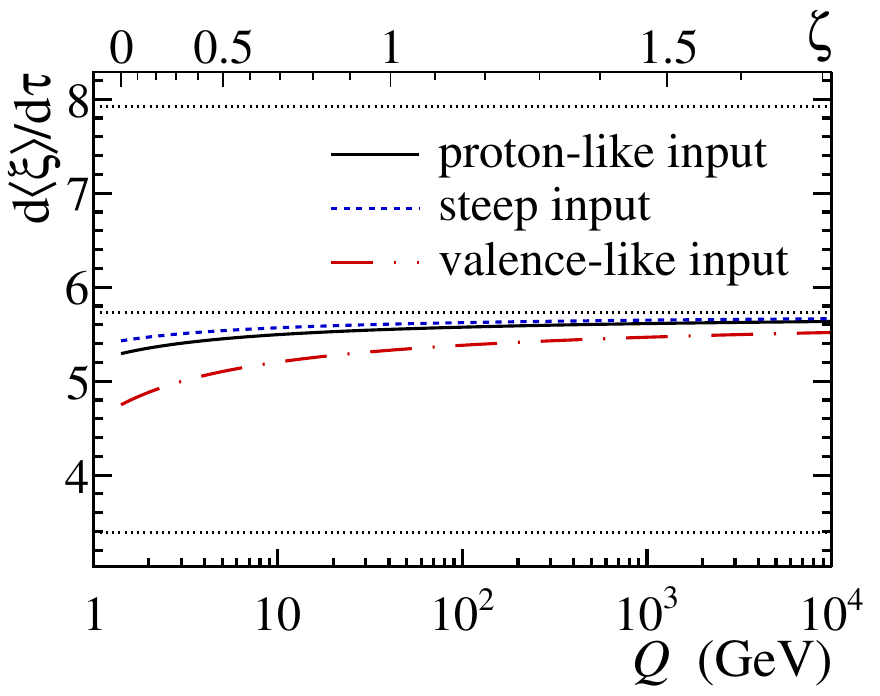}
\caption{Mean logarithmic momentum drift per unit $\tau$ at LO. The legend identifies the black solid proton-like, blue dashed steep, and red dash-dotted valence-like inputs. The steep input has rising small-$x$ gluon and sea densities, while the valence-like input has no small-$x$ rise. All start at $Q_0^2=2$~GeV$^2$ with one-loop running, $\Lambda_{\rm QCD}=\NumLambdaQCD$~GeV, and $n_f=5$. Dotted lines mark the edges \NumDriftLo{} and \NumDriftHi{} of Eq.~\eqref{eq:driftinterval} and the limit \NumMuTau{} of Eq.~\eqref{eq:driftlimit}. If the LO moment assumptions hold, the drift relaxes at rate $31/9$ per unit $\tau$, as in Eq.~\eqref{eq:xqrelax}. The drift stays between the edges along a forward path if and only if the input quark momentum fraction lies in $[0,1]$. This condition holds for the three inputs shown. The lower axis is logarithmic in $Q$, and the upper axis gives $\zeta$. The plotting program was written with AI assistance (Appendix~\ref{app:ai}).}
\label{fig:drift}
\end{figure*}

The quark momentum fraction follows the standard $N=2$ relaxation for the same five-flavor evolution, as given by Ellis, Stirling and Webber in Eqs.~(4.141)--(4.145) of Ref.~\cite{ESW1996},
\begin{equation}
\langle x\rangle_q(\tau)=\frac{15}{31}+\left(\langle x\rangle_q(0)-\frac{15}{31}\right)e^{-31\tau/9}.
\label{eq:xqrelax}
\end{equation}
Thus the drift relaxes monotonically at the rate $31/9$ per unit $\tau$ to
\begin{equation}
\lim_{\tau\to\infty}\frac{d\langle\xi\rangle}{d\tau}
=\frac{204\pi^2-415}{279}=2\pi b_0\,\mu=\NumMuTau .
\label{eq:driftlimit}
\end{equation}
The rate is $\mu=\NumMu$ per unit $\zeta$. Figure~\ref{fig:drift} shows the relaxation for the proton-like, steep, and valence-like inputs. The limiting species fractions are stationary, but splittings continue to move the mean logarithmic momentum. The zero eigenvalue conserves total momentum, and the nonzero eigenvalue fixes the relaxation of the species fractions.

\begin{figure*}[t]
\centering
\includegraphics[width=0.66\textwidth]{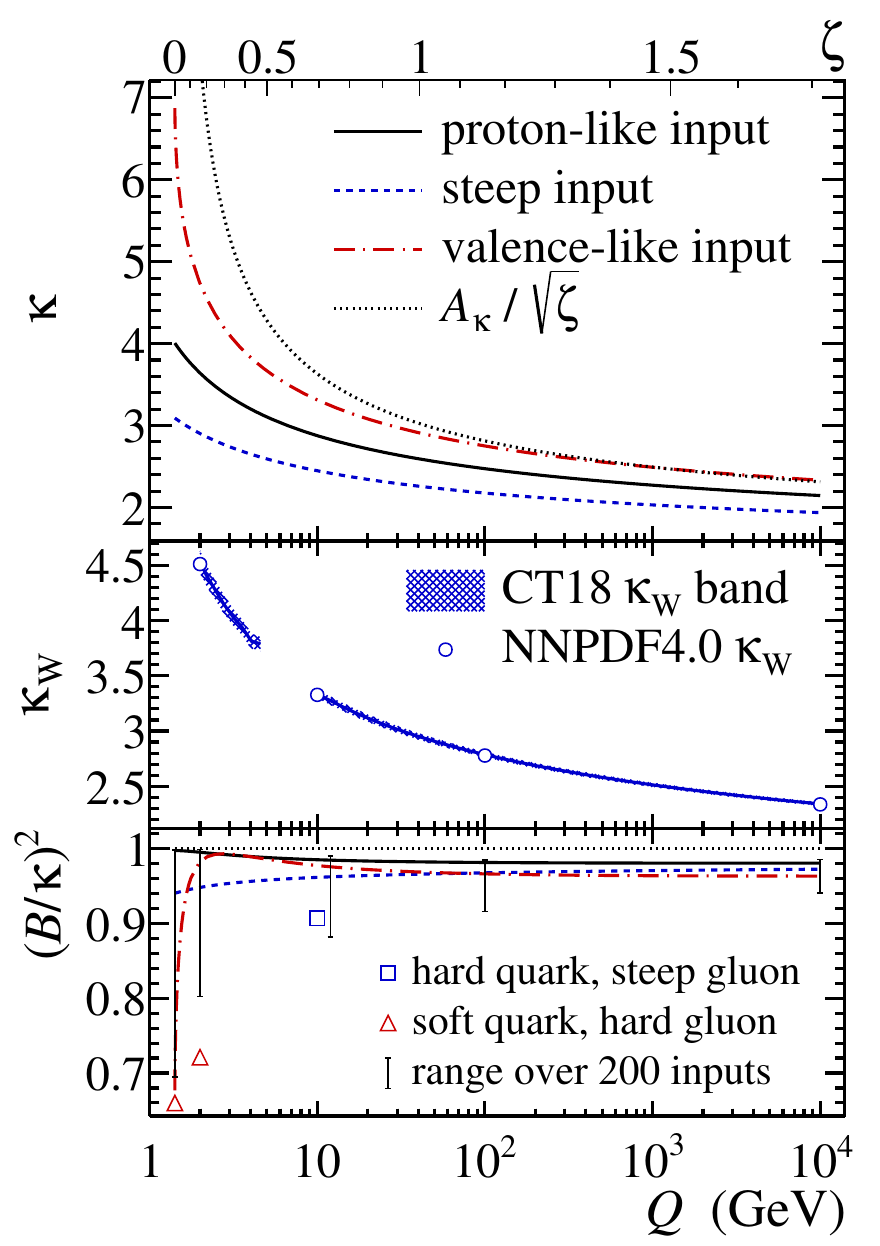}
\caption{Fisher speeds and squared mean-bound ratios. Upper-panel black solid, blue dashed, and red dash-dotted curves give full-interval LO $\kappa$ for the proton-like, steep, and valence-like inputs of Appendix~\ref{app:methods}. They start at $Q_0^2=2$~GeV$^2$ with one-loop running, $\Lambda_{\rm QCD}=\NumLambdaQCD$~GeV, and fixed $n_f=5$, including above the top threshold. The black dotted curve is the large-$\zeta$ reference $A_\kappa/\sqrt\zeta$, with $A_\kappa=\NumAkappa$ from Eq.~\eqref{eq:bounds}. The upper axis gives $\zeta$. The middle panel gives the windowed speed $\kappa_W$ of fitted NNLO densities on $W$ of Eq.~\eqref{eq:fitwindow}. The blue hatched CT18 band~\cite{CT18} has a blue solid central line and 68 percent parton-distribution and coupling uncertainty. It is absent from \NumBandFirstEnd{} to 10~GeV around and above $m_b=\NumMbCT$~GeV. Open blue circles give the NNPDF4.0 central member~\cite{NNPDF40} at the four scales of Table~\ref{tab:fitted}. These fitted densities, differentiated at their own order and coupling, are not the LO model curves. Appendix~\ref{app:methods} gives the CT18~\cite{CT18} scale-scan and band settings. Lower-panel LO curves give $(B/\kappa)^2$, with $B$ from Eq.~\eqref{eq:Bzeta}, and the dotted line marks unity. The open blue square is $\NumCorrSqHardqSteepg$ for the hard-quark, steep-gluon input at 10~GeV. Open red triangles give $\NumCorrSqSoftqHardg$ and $\NumCorrSqSoftqHardgQZero$ for the soft-quark, hard-gluon input at 2~GeV and $Q_0$. Black capped bars give the range of $(B/\kappa)^2$ over the \NumScanN{} scanned inputs of Sec.~\ref{sec:bound}~A at $Q_0$, 2, 10, 100, and $10^4$~GeV. The 10~GeV bar is drawn at 12~GeV beside the open square. The plotting program was written with AI assistance (Appendix~\ref{app:ai}).}
\label{fig:bound}
\end{figure*}

\subsection{Width and species memory}

Momentum exchanged between quarks and gluons with different mean logarithmic momenta changes the total variance. Appendix~\ref{app:cumulants} retains their separation in $C_q$, the quark contribution to the first moment centered on the total mean. The momentum-fraction solution gives the mean at every scale for $n_f=5$ with one-loop running,
\begin{equation}
\begin{gathered}
\langle\xi\rangle=\langle\xi\rangle_0+\mu\zeta+\eta\,(E-1),\\
E=e^{-\omega\zeta},\qquad
\omega=\frac{62}{69},\\
\eta=\frac{10\pi^2-17}{62}\left(\langle x\rangle_q(Q_0)-\frac{15}{31}\right).
\end{gathered}
\label{eq:meanxi}
\end{equation}
The evolution of $C_q$, together with the momentum fractions, gives the variance at every scale,
\begin{equation}
\operatorname{Var}_P(\xi)=D\zeta+V_\infty+\bigl(V_E+V_{\zeta E}\zeta\bigr)E-\eta^2E^2 .
\label{eq:varxi}
\end{equation}
The coefficients $V_\infty$, $V_E$, and $V_{\zeta E}$ are given in Eq.~\eqref{eq:varcoeffs}. They depend on the input quark momentum fraction, the centered quark moment $C_{q0}$, and the input variance. The initial mean sets the origin of the mean trajectory and drops out of the variance and speed bound.

The terms proportional to $\zeta$ describe persistent drift and broadening. The factors $E$ describe relaxation of the input and species composition, which a calculation keeping only the larger eigenvalue would miss. These finite-scale terms decay at large time, leaving drift and variance rates $\mu$ and $D$. Finite input logarithmic variance suffices for these limits. The species separation also contributes to the limiting variance rate (Sec.~\ref{sm:approach} of the Supplemental Material).

The column derivatives still give the momentum-weighted drift at next-to-leading order (NLO), with the two-loop correction included. These derivatives follow from the two-loop splitting functions of Curci, Furmanski and Petronzio~\cite{CurciFurmanskiPetronzio1980} and of Furmanski and Petronzio~\cite{FurmanskiPetronzio1980} in the PEGASUS transcription of Vogt~\cite{Vogt2005}. Appendix~\ref{app:nlo} gives the assumptions and the residuals on CT18~\cite{CT18}, which at 10 and 100~GeV resolve the two-loop term but not the three-loop term. The general bound of Eq.~\eqref{eq:identity} needs no LO kernel, but its closed form in Eq.~\eqref{eq:Bzeta} uses the LO variance, which is not extended to NLO here.

\section{The lower bound and its large-scale form}
\label{sec:bound}

A moving mean requires the density to change, with less relative change needed for a broad distribution than for a narrow one. The Cauchy-Schwarz inequality in the Fisher-Rao metric expresses this relation. Ito and Dechant~\cite{ItoDechant2020} and Nicholson, Garc\'{\i}a-Pintos, del Campo and Green~\cite{Nicholson2020} used the same general inequality in stochastic thermodynamics.

If the mean-bound assumptions of Sec.~\ref{sec:density} hold, the mean drift is
\begin{equation}
\frac{d\langle\xi\rangle}{d\tau}=\operatorname{Cov}_P(s_\tau,\xi),
\qquad
s_\tau=\d_\tau\ln P .
\label{eq:covdrift}
\end{equation}
Cauchy-Schwarz then gives
\begin{equation}
\kappa\,\bigl|\operatorname{corr}_P(s,\xi)\bigr|=B\equiv\frac{|d\langle\xi\rangle/d\tau|}{\operatorname{sd}_P(\xi)},
\qquad
\kappa\ge B .
\label{eq:identity}
\end{equation}
The correlation form requires positive information. In this display $s$ keeps its definition as the $t$ slope. Its correlation with $\xi$ is the same as that of $s_\tau$, since the slopes differ by a positive factor at each scale.

If the LO moment assumptions hold and the variance of $\xi$ is positive at the scale, the right-hand side has a closed form at $n_f=5$ with one-loop running,
\begin{equation}
B(\zeta)=2\pi b_0\,\frac{\bigl|\mu-\omega\,\eta\,E\bigr|}{\sqrt{D\zeta+V_\infty+(V_E+V_{\zeta E}\zeta)E-\eta^2E^2}} .
\label{eq:Bzeta}
\end{equation}
The bound $B$ as a function of $\zeta$ needs only the input quark momentum fraction, $C_{q0}$, and input variance. The Fisher integral instead needs the density and its derivative at every $x$. The closed form gives a lower bound from moments and therefore a check on any evaluation of the Fisher integral. The LO positivity assumption keeps the density nonnegative at every scale of an LO path, so Eq.~\eqref{eq:Bzeta} bounds $\kappa$ there if the other mean-bound assumptions hold.

Equality at a scale requires the centered slope to be proportional to the centered logarithmic momentum. A component orthogonal to $\xi-\langle\xi\rangle$ changes the shape without moving the mean and raises $\kappa$ without raising $B$. Near equality, which requires no Gaussian density, means that this orthogonal component is small in momentum-weighted mean square compared with $\kappa^2$. Appendix~\ref{app:bound} gives the equality condition and differentiation argument.

The squared ratio $(B/\kappa)^2$ ranges from $\NumCorrSqMin$ to $\NumCorrSqMax$ for the proton-like, steep, and valence-like inputs at 2, 10, 100, and $10^4$~GeV (Fig.~\ref{fig:bound}). A Gaussian with the same drift and width gives $\NumGaussCorrSqMin$ to $\NumGaussCorrSqMax$, so this replacement does not explain the near equality.

The Fisher information is $I_\zeta=\langle\xi\rangle'^2/V+V'^2/(2V^2)$ for a Gaussian in $\xi$ whose mean $\langle\xi\rangle$ and variance $V=\operatorname{sd}_P(\xi)^2$ follow those of $P$, with primes denoting $\zeta$ derivatives. Thus $(B/\kappa)^2=\bigl[1+V'^2/(2V\langle\xi\rangle'^2)\bigr]^{-1}$ equals $1/R_{\rm low}$ of Eq.~\eqref{eq:Rlow} at $\chi_3=\chi_4=0$, where $K_4=2\chi_2^2$. Here $R$ is the squared speed divided by the squared mean bound, Eq.~\eqref{eq:Rdef} of Appendix~\ref{app:bound}. For the Gaussian $R=R_{\rm low}$ because the Gaussian score is quadratic in $\xi$ ($I_\zeta=J_2$).

A slope affine in $\xi$ is produced by multiplying the current density by a power of $1/x$ and renormalizing over an infinitesimal scale step. The LO logarithmic slope is close to this affine form in momentum-weighted mean square for the three model inputs at the four scales above. Their positive drift makes the corresponding power favor small $x$. The LO slope of these inputs is closer still to an affine function of $\sqrt{\xi}$ at 10, 100 and $10^4$~GeV, as shown below.

A trial slope in place of $s$ in the correlation form of Eq.~\eqref{eq:identity} gives the $(B/\kappa)^2$ that the same density would have with that centered slope. A trial slope $\sqrt{\xi}$ gives squared momentum-weighted correlations with $\xi$ of $\NumNullSqrtXiMin$ to $\NumNullSqrtXiMax$ for the same evolved densities at the four scales, within the range $\NumCorrSqMin$ to $\NumCorrSqMax$ of the LO slope $s$. The $\sqrt{\xi}$ dependence is the small-$x$ form of the slope in double-asymptotic scaling~\cite{DeRujula1974,BallForte1994}. The slopes $\xi^2$ and $\ln\xi$, which curve more where most of the momentum lies, give $\NumNullXiSqMin$ to $\NumNullXiSqMax$ and $\NumNullLnXiMin$ to $\NumNullLnXiMax$, respectively, below the smallest value found for $s$. The slope $\ln(1-x)$, concentrated at large $x$, gives the much lower range $\NumNullLnOneMinusXMin$ to $\NumNullLnOneMinusXMax$. A value of $(B/\kappa)^2$ in the range found for $s$ therefore does not by itself single out a power reweighting.

The LO slope $s$ is closer to an affine function of $\sqrt{\xi}$ than to one of $\xi$ in momentum-weighted mean square for the proton-like, steep and valence-like inputs at 10, 100 and $10^4$~GeV.
Its squared correlation with $\sqrt{\xi}$ is \NumSlopeSqrtXiHighMin{} to \NumSlopeSqrtXiHighMax{} there, and with $\xi$ it is \NumSlopeXiHighMin{} to \NumSlopeXiHighMax{}, which are the values of $(B/\kappa)^2$.
One minus the squared momentum-weighted correlation of $s$ with $\sqrt{\xi}$ or $\xi$ is the momentum-weighted mean square of the residual of the best affine fit in that variable divided by the momentum-weighted mean square of $s$, as in Eq.~\eqref{eq:residual} for $\xi$.
The squared correlation with $\sqrt{\xi}$ exceeds that with $\xi$ in \NumSlopeSqrtBetterCount{} of the 12 cases, the three inputs at the four scales, with the proton-like and valence-like inputs at 2~GeV as the exceptions.
The power $p$ that maximizes the squared correlation of $s$ with $\xi^p$ over $\NumSlopePScanLo\le p\le\NumSlopePScanHi$ lies between \NumSlopePStarHighMin{} and \NumSlopePStarHighMax{} for the three inputs at 10, 100 and $10^4$~GeV.
A slope this close to an affine function of $\sqrt{\xi}$ is also close to an affine function of $\xi$ in momentum-weighted mean square, because $\sqrt{\xi}$ and $\xi$ are strongly correlated under $P$, as the trial slope above shows, so $(B/\kappa)^2$ stays near one at these scales.
The comparison weights every $x$ by its momentum, and large $x$ contributes as well as small $x$, so it does not test the small-$x$ form of double-asymptotic scaling.

The ratio depends on input and scale. The two other model inputs give $\NumCorrSqExtraMin$ to $\NumCorrSqExtraMax$ at 2, 10, 100 and $10^4$~GeV. At $Q_0$, the steep input gives $\NumCorrSqSteepQZero$ and the valence-like input $\NumCorrSqDenseMin$. Appendix~\ref{app:methods} gives the complete evaluated ranges for the hard-quark, steep-gluon and soft-quark, hard-gluon inputs. At 2~GeV the soft-quark, hard-gluon value $\NumCorrSqSoftqHardg$ lies below its Gaussian value $\NumGaussSoftqHardg$. The LO trajectory from CT18~\cite{CT18} gives $\NumCorrSqCTLOMin$ to $\NumCorrSqCTLOMax$ at 2, 5, 10, 100, and $10^4$~GeV.

\subsection{Effective power and the residual slope}

The part of the scale slope that moves the mean is affine in $\xi$. The scale slope decomposes orthogonally, if the mean-bound assumptions hold and $0<\kappa<\infty$, as
\begin{equation}
\begin{gathered}
s_\tau=\beta\,\bigl(\xi-\langle\xi\rangle\bigr)+r,\qquad
\beta=\frac{1}{\operatorname{Var}_P(\xi)}\,\frac{d\langle\xi\rangle}{d\tau},\\
\langle r\rangle_P=\operatorname{Cov}_P(r,\xi)=0,\qquad
\frac{B^2}{\kappa^2}=1-\frac{\langle r^2\rangle_P}{\kappa^2}.
\end{gathered}
\label{eq:residual}
\end{equation}
Here $r$ is the residual and $\beta$ the effective-power rate. The deficit $1-(B/\kappa)^2$ is the residual mean square, weighted by momentum, relative to $\kappa^2$.

The affine term is the slope of normalized reweighting by a power of $1/x$. If $P=x^{-\lambda_0(\tau)}g(x)$ divided by its normalization, with $g$ fixed, then $r=0$ and $\beta=d\lambda_0/d\tau$. For a general density, $\beta$ averages the rate of the local exponent in Eq.~\eqref{eq:lambdadef} with the weight $\mathcal T$,
\begin{equation}
\begin{gathered}
\beta=\frac{1}{\operatorname{Var}_P(\xi)}\int_0^\infty d\xi\,\mathcal T(\xi)\,\d_\tau\lambda,\\
\mathcal T(\xi)=\int_\xi^\infty dy\,\bigl(y-\langle\xi\rangle\bigr)\,p(y)\ge0,\qquad
\int_0^\infty d\xi\,\mathcal T(\xi)=\operatorname{Var}_P(\xi).
\end{gathered}
\label{eq:betaweight}
\end{equation}
Here $p=p_q+p_g$ is the density in $\xi$, and $\d_\tau\lambda$ is taken at fixed $\xi$. This identity assumes $P>0$ on $0<x<1$, $\ln P$ twice continuously differentiable in $\xi$ and $\tau$, $s_\tau\mathcal T$ vanishing at both ends of the $\xi$ range, and finite $\int_0^\infty|\d_\tau\lambda|\mathcal T\,d\xi$. Appendix~\ref{app:path} derives it. The local rate need not equal $\beta$ and, for the proton-like input at 2~GeV, is $\NumLamRateRatioA\,\beta$ at $x=10^{-6}$.

Equation~\eqref{eq:powerwindow} gives the normalized density for $F_2=c(Q^2)x^{-\lambda(Q^2)}$ on a fixed window. The slope in Eq.~\eqref{eq:windowslope} is affine in $\ln(1/x)$, so the windowed mean bound $B_W=|d\langle\xi\rangle_{P_W}/d\ln Q^2|/\operatorname{sd}_{P_W}(\xi)$ equals $v_W$ of Eq.~\eqref{eq:windowspeed}, whether or not $c$ depends on $Q^2$. The bound $B_W$ is the analog of $B$ per unit $\ln Q^2$. The H1 description~\cite{H1rise2001} has this single-power form, with $d\lambda/d\ln Q^2=\NumHOneRate$ at $x\le10^{-2}$ and $Q^2\ge\NumHOneQSqMin$~GeV$^2$. Any single power reaches the bound, whatever its exponent, so the equality $B_W=v_W$ for the H1 description gives no result about the H1 data. Section~\ref{sm:heraresults} of the Supplemental Material compares the windowed speed of the normalized structure function from the combined H1 and ZEUS data~\cite{HERAcomb} with the expectation from the H1 power-law rate. For $P$, near equality means that its LO logarithmic rate of change is close in momentum-weighted mean square to a slope affine in $\ln(1/x)$. The residual $r$ measures the difference.

For the proton-like input, much of $\langle r^2\rangle_P$ comes from the ends of the momentum range. At 2~GeV, $x<\NumRegEdgeSmall$ contributes \NumRegDefSmallTwoA{} percent with \NumRegMomSmallTwoA{} percent of the momentum. At $10^4$~GeV, $x>\NumRegEdgeLarge$ contributes \NumRegDefLargeTenTeVA{} percent with \NumRegMomLargeTenTeVA{} percent of the momentum. Near $x=1$, the quark-dominated slope behaves as $2C_F\ln(1-x)$, as in Eq.~\eqref{eq:xtoone}, and is not affine in $\xi$.

The input scan uses \NumScanN{} LO inputs with each species given by one term $c\,\mathcal B_{a,b}$ of Appendix~\ref{app:methods}, with fixed $n_f=5$, one-loop running, $Q_0^2=2$~GeV$^2$, and $\Lambda_{\rm QCD}=\NumLambdaQCD$~GeV. Section~\ref{sm:scan} of the Supplemental Material defines the grid and tabulates ranges and medians. Over these inputs, $(B/\kappa)^2$ ranges from $\NumScanMinQZero$ to $\NumScanMaxQZero$ at $Q_0$, $\NumScanMinTwo$ to $\NumScanMaxTwo$ at 2~GeV, and $\NumScanMinTenTeV$ to $\NumScanMaxTenTeV$ at $10^4$~GeV (Fig.~\ref{fig:bound}, lower panel). At 10~GeV, \NumScanCountNinetyTen{} inputs reach or exceed $0.9$, and all do at 100 and $10^4$~GeV. From $Q_0$ to $10^4$~GeV the minimum rises from \NumScanMinQZero{} to \NumScanMinTenTeV{}, and the maximum at $10^4$~GeV is below its value at $Q_0$, so the range narrows toward \NumScanMinTenTeV{} to \NumScanMaxTenTeV{} and not toward equality. No predictive formula for $r$ is known.

The input giving the minimum at $Q_0$ has quark and gluon densities that both vanish at small $x$. The pole at $N=1$ of the LO matrix's gluon row then makes $\d_\tau P$ tend to a nonzero constant at small $x$ while the input density vanishes. The resulting large slope is reduced within a short evolution interval. For the valence-like input, $(B/\kappa)^2$ rises from $\NumCorrSqDenseMin$ at $Q_0$ to $\NumCorrSqValenceTwo$ at 2~GeV. Its deficit halves at $\tau=\NumTransTauHalf$, or $Q^2=\NumTransQsqHalf$~GeV$^2$, \NumTransRatio{} times sooner than the quark-fraction relaxation time $9/31$ of Eq.~\eqref{eq:xqrelax}. Appendix~\ref{app:path} derives the leading small-$\tau$ law.

\subsection{Large evolution time}

At large time the finite input moments no longer affect the leading coefficient. The large-time limit for any positive coupling for which $\tau$ tends to infinity is
\begin{equation}
\lim_{\tau\to\infty}\sqrt{\tau}\,B=-\frac{\gamma_+^{(0)\prime}(2)}{\sqrt{\gamma_+^{(0)\prime\prime}(2)}}=\NumTauConst .
\label{eq:tauconst}
\end{equation}
One-loop running gives
\begin{equation}
\begin{gathered}
\liminf_{\zeta\to\infty}\sqrt{\zeta}\,\kappa\ge A_\kappa=\frac{2\pi b_0\,\mu}{\sqrt D}=\NumAkappa,\\
\liminf_{\zeta\to\infty}\frac{L}{\sqrt\zeta}\ge A_L=\frac{2\mu}{\sqrt D}=\NumAL .
\end{gathered}
\label{eq:bounds}
\end{equation}
Here $\kappa$ is the speed per unit $\tau$, while the limit is taken in $\zeta$. The values are for $n_f=5$, with other flavor numbers in Appendix~\ref{app:matrix}. The length starts at a fixed scale. The finite moment equations give $A_\kappa$ and $A_L$, and the two limit-inferior bounds also require $\kappa\ge B$ at all sufficiently large $\zeta$. The same drift-and-broadening mechanism appears in the Fisher information of biased Brownian motion, a Gaussian whose mean moves at constant speed while its variance grows linearly (Ref.~\cite{ItoDechant2020}, Sec.~VI~B, Eq.~(79)). We derive its LO DGLAP coefficients.

These limit-inferior bounds give no inequality at a specified finite scale. For the proton-like input, $A_\kappa/\sqrt\zeta$ exceeds the computed speed by factors $\NumAsymOverKappaTwo$ and $\NumAsymOverKappaTenTeV$ at 2 and $10^4$~GeV. The valence-like speed crosses the reference from below before the highest plotted scale. At finite scale the relevant lower bound is Eq.~\eqref{eq:Bzeta}.

The quadratic-moment assumptions also let moments of $\ln(1/x)$ through fourth order bound how fast $(\kappa/B)^2$ can approach unity at large $\zeta$. Evaluations at $\zeta$ of order a hundred for the inputs tested in Sec.~\ref{sm:approach} of the Supplemental Material are consistent with the asymptotic coefficient. These times correspond to no physical scale. At the plotted scales, $\zeta(R-1)$ is still rising, so these scales do not test the coefficient.

\subsection{Length and distance}

Wootters~\cite{Wootters1981} defined statistical distance for finitely or countably many outcomes as the statistical length of the shortest connecting curve and, with the amplitudes $\sqrt{p_i}$ on the unit sphere, found it to be the angle $\arccos\sum_i\sqrt{p_ip_i'}$ (Eqs.~(10) and~(11) of Ref.~\cite{Wootters1981}). That statistical length is $L/2$ in our normalization. Normalized densities on $0<x<1$ with measure $dx$ map by $u=2\sqrt P$ to the sphere of radius 2 in the square-integrable functions on this interval, with metric $ds^2=I_F\,dt^2$ (Eq.~\eqref{eq:sphere}). If the Fisher assumptions hold and $u$ is absolutely continuous in $t$ in this space,
\begin{equation}
L(t_1,t_2)\ge D_{\rm FR}(t_1,t_2)=2\arccos\int_0^1 dx\,\sqrt{P(x,t_1)\,P(x,t_2)} .
\label{eq:dfr}
\end{equation}
Equality requires the great-circle arc of Eq.~\eqref{eq:geodesic} traversed without reversal, so $D_{\rm FR}$ is the shortest path length between the densities. Appendix~\ref{app:path} derives the inequality and this condition. The distance needs only the end densities, and the second lower bound, $\int B\,d\tau$ in Eq.~\eqref{eq:lengthbound}, only the three input numbers.

The length $L=\NumGeoLA$ exceeds $D_{\rm FR}=\NumGeoDA$ by \NumGeoExcessA{} percent and $\int B\,d\tau=\NumGeoBA$ by \NumGeoExcessBA{} percent for the proton-like LO path from $Q_0$ to $10^4$~GeV. The length $L$ exceeds $D_{\rm FR}$ over the five model inputs by \NumGeoExcessMin{} to \NumGeoExcessMax{} percent from $Q_0$ and \NumGeoExcessTwoMin{} to \NumGeoExcessTwoMax{} percent from 2~GeV. The LO paths are therefore close in length to the shortest paths between their end densities. From $Q_0$ to $10^4$~GeV, $L$ exceeds the larger of its two lower bounds by at most \NumGeoBestExcessMax{} percent for each of the five model inputs (Sec.~\ref{sm:scan} of the Supplemental Material).

At small separation, the excess of $L$ over $D_{\rm FR}$ is set by the geodesic curvature $k$ at the path midpoint,
\begin{equation}
\frac{L(t_1,t_2)}{D_{\rm FR}(t_1,t_2)}-1=\frac{k^2L^2}{24}+O\bigl(L^4\bigr).
\label{eq:curvlaw}
\end{equation}
The expansion assumes four derivatives of $u$ in $t$ in the square-integrable functions, $I_F>0$, and finite $\langle w^2\rangle_P$ in Eq.~\eqref{eq:curvature}. These moments diverge for the valence-like input at $Q_0$, so the expansion does not apply. For the proton-like input, $k^2$ of the LO path is \NumGeoCurvLOTwoA{} at 2~GeV and \NumGeoCurvLOTenTeVA{} at $10^4$~GeV, below \NumGeoCurvTiltTwoA{} and \NumGeoCurvTiltTenTeVA{} for the reweighting path through the same density, Eq.~\eqref{eq:tiltlength} (Sec.~\ref{sm:scan} of the Supplemental Material).

We follow the reweighting path of Eq.~\eqref{eq:tiltpath} from $P(Q_0)$ to the distance $D_{\rm FR}$ that the LO path reaches at $10^4$~GeV for each of the five model inputs. This reweighting path is longer than the LO path. It exceeds $D_{\rm FR}$ by \NumGeoTiltExcessA{} percent for the proton-like input and \NumGeoTiltExcessMin{} to \NumGeoTiltExcessMax{} percent over the five inputs. At $10^4$~GeV the proton-like and steep densities lie at distances \NumGeoTiltDistA{} and \NumGeoTiltDistSteep{} from the whole reweighting families through their $P(Q_0)$. Closeness of $\kappa$ to $B$ is a property of the tangent at each scale and does not make the evolved density a reweighting of the input.

\section{The Markov structure, the upper bound, windows, and thresholds}
\label{sec:markov}

The species label gives information beyond the momentum fraction alone. The sum of the quark and gluon densities discards that distinction and can reduce the Fisher speed. At LO, the joint distribution evolves by splittings that increase $\xi$ and may change the species~\cite{JadachSkrzypek2004,Teryaev1998}.

The positive, time-homogeneous evolution in $\tau$ has the form
\begin{equation}
\begin{gathered}
p_j(\cdot,\tau)=\sum_{k=q,g}U_{jk}(\tau)\ast p_k(\cdot,0),\\
U_{jk}(\tau,d\ell)\ge0\ \text{on}\ \ell\ge0,\\
\int_0^\infty e^{-(N-2)\ell}\,U_{jk}(\tau,d\ell)=\bigl[e^{\tau\Gamma(N)}\bigr]_{jk},\\
\sum_{j=q,g}\int_0^\infty U_{jk}(\tau,d\ell)=1 .
\end{gathered}
\label{eq:markov}
\end{equation}
Here $U_{jk}(\tau,d\ell)$ is the probability that momentum initially in species $k$ reaches species $j$ with its value of $\ln(1/x)$ increased by an amount in $d\ell$. Column normalization expresses momentum conservation. Appendix~\ref{app:markov} gives the gain and loss terms and the connection to multitype fragmentation~\cite{Bertoin2008}. The species sum generally has no closed scalar Markov operator because its derivative depends on the quark-gluon division of momentum at each $x$.

If the LO positivity assumption of Sec.~\ref{sec:density} holds and the joint information is finite, the information lost on summing species is
\begin{equation}
\begin{gathered}
\kappa_{\rm joint}^2-\kappa^2=\int_0^1 dx\,\frac{P_q\,P_g}{P_q+P_g}\bigl(s_{\tau,q}-s_{\tau,g}\bigr)^2\ge0,\\
\kappa_{\rm joint}^2=\sum_{j=q,g}\int_0^1 dx\,\frac{(\d_\tau P_j)^2}{P_j},\qquad
s_{\tau,j}=\d_\tau\ln P_j .
\end{gathered}
\label{eq:jointgap}
\end{equation}
The gap depends on the difference of the species slopes where both species have nonzero momentum density, and vanishes if those slopes agree there. Equation~\eqref{eq:jointgap} holds at any value of $(B/\kappa)^2$. The ratio $\kappa_{\rm joint}/\kappa$ lies between $\NumKjointRatioMin$ for the valence-like input at 10~GeV and $\NumKjointRatioMax$ for the proton-like input at 2~GeV, over the three plotted inputs at 2, 10, 100, and $10^4$~GeV.

At $Q_0$, the hard-quark, steep-gluon input of Appendix~\ref{app:methods} has $P_q$ vanishing like $x^{\NumInHqSgQa}$ while $\partial_\tau P_q$, fed by the gluon, grows like $x^{\NumInHqSgGa}$, so its joint information is infinite and the requirement of finite joint information in the Markov-contraction assumptions and Eq.~\eqref{eq:jointgap} fails there. At $Q_0$, the soft-quark, hard-gluon input has $P_g$ vanishing like $x^{\NumInSqHgGa}$ while $\partial_\tau P_g$ grows like $x^{\NumInSqHgQa}$, with the same infinite joint information and failure of that requirement.

Fisher information contracts under a common Markov transition, as shown by Ay, Jost, L\^e and Schwachh\"ofer~\cite{AJLS2015}. For neighboring times, the Markov-contraction assumptions of Sec.~\ref{sec:density} give
\begin{equation}
\begin{gathered}
\kappa_{\rm joint}(\tau_2)\le\kappa_{\rm joint}(\tau_1),\\
\tau_1\le\tau_2\ \text{in one flavor segment}.
\end{gathered}
\label{eq:jointmono}
\end{equation}
Ito and Dechant~\cite{ItoDechant2020} prove the same monotonicity for time-independent Markov jump and Fokker-Planck dynamics by differentiating the Fisher information (Eq.~(15) and Appendix~D, Eq.~(D17), of Ref.~\cite{ItoDechant2020}). Appendix~\ref{app:markov} gives the continuum step. The generator must be independent of evolution time, as it is in $\tau$ at fixed $n_f$, but not in $t$ with a running coupling. The result applies to the joint LO density within a fixed-flavor segment. It implies neither monotonicity of the species sum nor ordering across a threshold where the generator changes. The joint speed per unit $\ln Q^2$, $(\alpha_s/2\pi)\,\kappa_{\rm joint}$, is also non-increasing within the segment wherever $\alpha_s$ decreases with $Q$, as for one-loop running.

\subsection{Restriction to a window}

A window discards momentum fractions and renormalizes what remains. The Fisher assumptions and differentiation under the integral that defines $Z_W$, $\d_\tau Z_W=\int_W\d_\tau P\,dx$, give the window split for fixed $W$ with $0<Z_W<1$,
\begin{equation}
\begin{gathered}
I_\tau=\frac{\bigl(\d_\tau Z_W\bigr)^2}{Z_W(1-Z_W)}+Z_W\,I_W+(1-Z_W)\,I_{W^c},\\
Z_W=\int_W P\,dx .
\end{gathered}
\label{eq:windowsplit}
\end{equation}
Here $I_W=\kappa_W^2$ and $I_{W^c}$ refer to the separately normalized restrictions. The first term measures the changing momentum fraction in $W$, and the other terms measure shape changes inside and outside it. Nonnegativity gives the left inequality below. The right follows from Eq.~\eqref{eq:jointgap} with the LO positivity assumption and finite joint information,
\begin{equation}
Z_W\,\kappa_W^2\le\kappa^2\le\kappa_{\rm joint}^2,
\qquad
\kappa_W\le\frac{\kappa}{\sqrt{Z_W}} .
\label{eq:window}
\end{equation}
Normalization of the retained density can increase its speed, so the factor $Z_W$ is necessary. A change of momentum fraction in the window contributes to the full-interval information even when normalization removes it from the windowed shape.

The full-interval lower bound $B$ does not bound $\kappa_W$. The windowed speed $\kappa_W$ on the window of Eq.~\eqref{eq:fitwindow} is $\NumKappaWTenTeVA$, below $B=\NumBTenTeVA$, for the proton-like input at $10^4$~GeV. The windowed speed lies below the full-interval quadratic-moment bound of Eq.~\eqref{eq:Jtwo}, converted to $\tau$ units, by $\NumWinBelowJTwoHundred$ percent at 100~GeV and $\NumWinBelowJTwoTenTeV$ percent at $10^4$~GeV on the LO trajectory from CT18~\cite{CT18}. At 2, 5, and 10~GeV it lies above that bound by \NumWinAboveJTwoTwo, \NumWinAboveJTwoFive, and \NumWinAboveJTwoTen{} percent.

\subsection{Flavor thresholds}

At a flavor threshold, an additional splitting channel changes the generator. If the LO matching assumptions of Sec.~\ref{sec:density} hold,
\begin{equation}
\kappa_+^2-\kappa_-^2=2\operatorname{Cov}_P(s_\tau,\delta)+\operatorname{Var}_P(\delta),
\qquad
\delta=\frac{\Delta K}{P}.
\label{eq:thresholdjump}
\end{equation}
The slope $s_\tau$ is evaluated below the threshold, and $\Delta K$ is the change in total kernel action. The covariance has no prescribed sign. The joint monotonicity of Eq.~\eqref{eq:jointmono} applies within a fixed-flavor segment and does not fix the sign of $\kappa_+^2-\kappa_-^2$ in Eq.~\eqref{eq:thresholdjump}.

The jump in mean drift is fixed by the gluon momentum fraction,
\begin{equation}
\Delta\frac{d\langle\xi\rangle}{d\tau}=\operatorname{Cov}_P(\xi,\delta)=-2\,\Gamma_{hg}'(2)\,\langle x\rangle_g=\frac{11}{72}\,\langle x\rangle_g .
\label{eq:thresholddrift}
\end{equation}
Here $\Gamma_{hg}$ is defined in Eq.~\eqref{eq:dGamma}. The per-flavor change $11/72$ of the gluon rate also moves the upper edge of Eq.~\eqref{eq:driftinterval}. The drift jump is positive when the gluon momentum fraction is positive. The full-interval speed increases by $\NumMbStepLO$ percent at $m_b=\NumMbCT$~GeV on the LO trajectory from CT18~\cite{CT18}. Appendix~\ref{app:markov} also distinguishes this finite jump from the information singularity obtained when the newly opened heavy species is resolved separately.

\section{Fitted densities}
\label{sec:fitted}

Fitted densities change both the momentum distribution and its scale derivative. We compare their speeds with the LO results using CT18~\cite{CT18} and NNPDF4.0~\cite{NNPDF40}, and examine positivity of the MSHT20 global fit of Bailey et~al.~\cite{MSHT20} at NNLO. The grids from the LHAPDF library of Buckley et~al.~\cite{LHAPDF} are differentiated at each fit's own order and coupling convention. Each speed is divided by the $\alpha_s(Q)$ tabulated with that fit, with the convention dependence specified in Sec.~\ref{sec:density}~B.

The fixed normalization window is
\begin{equation}
W=\bigl[10^{-6},\,0.99\bigr].
\label{eq:fitwindow}
\end{equation}
The lower edge $10^{-6}$ is the first $x$ node of the MSHT20 grid~\cite{MSHT20} and $W$ is the same for the three global fits. We retain the normalization derivative after checking positivity throughout this window,
\begin{equation}
\d_t\ln P_W=\frac{\d_tP}{P}-\d_t\ln Z_W,
\qquad
P_W=\frac{P}{Z_W}.
\label{eq:windowcenter}
\end{equation}
We then evaluate
\begin{equation}
\begin{aligned}
\kappa_W^2&=\operatorname{Var}_{P_W}\!\left(\frac{K_{\rm tot}}{P}\right)\\
&=\frac{1}{Z_W}\int_W dx\,\frac{K_{\rm tot}^2}{P}-\left(\frac{1}{Z_W}\int_W dx\,K_{\rm tot}\right)^2 .
\end{aligned}
\label{eq:kappadef}
\end{equation}
Here $K_{\rm tot}$ is obtained from the fitted density's scale derivative. The variance removes the common normalization change.

The CT18 speed~\cite{CT18} falls by a factor $\NumFallTwoTenTeV$ between 2 and $10^4$~GeV (Table~\ref{tab:fitted}). The fraction $\NumFallShareBelowTen$ of this fall of the CT18 speed~\cite{CT18} occurs below 10~GeV. The NNPDF4.0 central speeds~\cite{NNPDF40} are below those of CT18~\cite{CT18} at all four scales, by \NumNNPDFCTMaxDiff{} percent at 2~GeV and less at higher scales. The CT18 uncertainty~\cite{CT18} describes its Hessian and coupling variations, including changes of the fitted shape. It excludes changes of normalization interval and the range of derivative prescriptions near thresholds. Section~\ref{sm:fitmethods} of the Supplemental Material gives these effects and the separate interpolation systematic. These fit uncertainties are not errors on the LO moment formulas.

\begin{table}[t]
\caption{Dimensionless windowed speed $\kappa_W$ per unit $\tau$ for the fitted central densities, using Eq.~\eqref{eq:fitwindow} and the normalization derivative. CT18~\cite{CT18} uncertainties combine parton-distribution and coupling variations at 68 percent confidence level. NNPDF4.0~\cite{NNPDF40} entries use its central member. The scale derivative is the bicubic-spline derivative of Sec.~\ref{sm:fitmethods} of the Supplemental Material at every scale, including 2~GeV below the bottom threshold. The final column gives the NNPDF4.0 speed~\cite{NNPDF40} divided by the CT18 central speed~\cite{CT18}, minus one, in percent, computed before rounding. Section~\ref{sec:fitted}~A gives the positive-replica spreads~\cite{NNPDF40}. Lowering $x_{\min}$ from $10^{-6}$ to $10^{-9}$ changes the CT18 speed~\cite{CT18} at $10^4$~GeV by \NumXminSensTenTeV{} percent, outside the quoted uncertainty (Sec.~\ref{sm:fitmethods} of the Supplemental Material).}
\label{tab:fitted}
\begin{ruledtabular}
\begin{tabular}{cccc}
$Q\;(\mathrm{GeV})$ & CT18~\cite{CT18} & NNPDF4.0~\cite{NNPDF40} & Difference (percent)\\
\hline
$2$ & $\NumKappaCTTwo\pm\NumKappaCTUncTwo$ & \NumKappaNNPDFTwo & \NumNNPDFCTRelTwo\\
$10$ & $\NumKappaCTTen\pm\NumKappaCTUncTen$ & \NumKappaNNPDFTen & \NumNNPDFCTRelTen\\
$100$ & $\NumKappaCTHundred\pm\NumKappaCTUncHundred$ & \NumKappaNNPDFHundred & \NumNNPDFCTRelHundred\\
$10^4$ & $\NumKappaCTTenTeV\pm\NumKappaCTUncTenTeV$ & \NumKappaNNPDFTenTeV & \NumNNPDFCTRelTenTeV\\
\end{tabular}
\end{ruledtabular}
\end{table}

The fitted speed need not be monotone at thresholds. The CT18 scan~\cite{CT18} decreases between them but rises near the bottom threshold for each of the three derivative prescriptions in Sec.~\ref{sm:fitmethods} of the Supplemental Material. The size of the rise depends on that prescription, so the region around and above the bottom threshold is omitted from the fitted band. The joint LO contraction of Eq.~\eqref{eq:jointmono} does not apply to this fitted NNLO scan.

At $Q=10$~GeV, $x<10^{-2}$ contains $\NumMomShareTen$ percent of the momentum in the window and $\NumFisherShareTen$ percent of its Fisher information for CT18~\cite{CT18}. Section~\ref{sm:fitmethods} of the Supplemental Material gives the smaller-$x$ shares, the reach of the combined H1 and ZEUS data~\cite{HERAcomb}, and the resulting dependence on the fitted input at small $x$.

The LO kernel applied to the same CT18 density~\cite{CT18} isolates the change in derivative. The LO trajectory from CT18~\cite{CT18} also changes the density as it evolves. Section~\ref{sm:fitmethods} of the Supplemental Material gives their speeds and separates the effects of the derivative and normalization window at 2 and $10^4$~GeV.

\subsection{Positivity and choice of window}

A fitted density defines a Fisher speed only where it is nonnegative throughout the normalization interval. The input gluon of MSHT20 may be negative at very small $x$ (footnote~5 of Ref.~\cite{MSHT20}). The total density is also negative on part of the window at low scales in the MSHT20 scans~\cite{MSHT20} of Sec.~\ref{sm:fitmethods} of the Supplemental Material, so its full-window speed is undefined there. NNPDF4.0~\cite{NNPDF40} imposes positivity at $Q^2=\NumNNPDFPosQsq$~GeV$^2$ (Eq.~(3.10) of Ref.~\cite{NNPDF40}), above 2~GeV. At 2~GeV, \NumNNPDFNusedTwo{} of the \NumNNPDFNtotTwo{} NNPDF4.0 replicas~\cite{NNPDF40} are positive throughout the window, with speed standard deviation $\NumNNPDFSpreadTwo$ percent of the central value. This spread applies only to the positive subset because the omitted replicas are negative somewhere in the interval.
At 10, 100, and $10^4$~GeV, the standard deviations are \NumNNPDFSpreadTen, \NumNNPDFSpreadHundred, and \NumNNPDFSpreadTenTeV{} percent of the central value, from \NumNNPDFNusedTen, \NumNNPDFNusedHundred, and \NumNNPDFNusedTenTeV{} replicas positive throughout the window, respectively.

\subsection{Other fits and measures}

The NNPDF3.1sx fits~\cite{BallResum2018} at NNLO without and with small-$x$ resummation give windowed speeds on $10^{-6}\le x\le\NumResWindowHi$. The resummed central speed divided by the fixed-order central speed, minus one, is $\NumResObsTwo$ percent at 2~GeV and $\NumResObsHundred$ percent at 100~GeV. Section~\ref{app:resummed} of the Supplemental Material separates the contributions of their fitted inputs and of their kernels.

The combined H1 and ZEUS data~\cite{HERAcomb} also permit a speed for the normalized structure function under the measure $F_2\,dx$, evaluated in Sec.~\ref{app:hera} of the Supplemental Material. It is a separate statistic of measured scaling violations, with no derived equality or inequality relating it to the momentum-density information. The pole model of Sec.~\ref{app:pole} of the Supplemental Material instead gives a solvable distribution of total momentum loss. Its drift and width follow from retaining the leading Mellin pole and replacing the regular part by a normalization constant. The full LO matrix gives $D/\mu=\NumDoverMu$ and must be used for the singlet bounds.

\section{Conclusions and discussion}
\label{sec:conclusions}

We have shown that the speed per unit evolution time has a closed-form lower bound $B$ in DGLAP evolution at leading order (LO) with five fixed flavors, one-loop running and splitting functions, separately nonnegative input densities for quarks and gluons, the mean-bound assumptions and LO moment assumptions in Table~\ref{tab:assumptions}, and a positive variance of $\xi$ at each scale of evaluation.
The bound at any evolution time follows from three input numbers and the coupling.
The closed forms of the drift and width retain how the input momentum is shared between quarks and gluons and need no evolved density.
The squared ratio of the bound $B$ to the speed $\kappa$ is $\NumCorrSqMin$ to $\NumCorrSqMax$ for three of the five model inputs, the proton-like, steep and valence-like inputs, evaluated at 2, 10, 100 and $10^4$~GeV.
The CT18 global fit at NNLO (next-to-next-to-leading order) of the CTEQ-TEA group~\cite{CT18} supplies a separate input for an LO trajectory that uses the fit's coupling and includes a bottom-quark threshold. The squared ratios on this trajectory are $\NumCorrSqCTLOMin$ to $\NumCorrSqCTLOMax$ at the five scales 2, 5, 10, 100 and $10^4$~GeV, and $B$ is computed from the evolved density since the closed form requires five flavors throughout.

Momentum conservation makes the nonnegative momentum-weighted density a probability distribution, so evolution traces a path through distributions.
We apply information geometry to LO DGLAP evolution in its probabilistic form. We calculate the momentum density's Fisher information and speed along $\ln Q^2$ for the first time, to our knowledge.
The speed is the momentum-weighted standard deviation of the logarithmic rate of change, per unit of the chosen evolution time.
The mean bound uses the Cauchy-Schwarz inequality in the form applied by Ito and Dechant~\cite{ItoDechant2020} and by Nicholson, Garc\'{\i}a-Pintos, del Campo and Green~\cite{Nicholson2020} to stochastic thermodynamics.
The new elements of the bound here are the closed-form width of $\ln(1/x)$ at every scale and the interval of its drift for coupled quark-gluon evolution.

Near equality at a scale means that the density's LO logarithmic rate of change is close to a slope affine in $\ln(1/x)$.
The closeness is in momentum-weighted mean square, not at each $x$.
The deficit $1-(B/\kappa)^2$ is the momentum-weighted mean square of the residual slope $r$, the part of the logarithmic slope that is not affine in $\ln(1/x)$, divided by $\kappa^2$.
A slope varying as $\sqrt{\ln(1/x)}$, the small-$x$ form of the slope in double-asymptotic scaling~\cite{DeRujula1974,BallForte1994}, would give $(B/\kappa)^2$, the squared momentum-weighted correlation of the slope with $\ln(1/x)$, of $\NumNullSqrtXiMin$ to $\NumNullSqrtXiMax$ for the same three inputs and scales, inside the range $\NumCorrSqMin$ to $\NumCorrSqMax$ found for LO evolution.
Near equality therefore does not by itself single out a power reweighting.
The LO slope is closer to an affine function of $\sqrt{\ln(1/x)}$ than to one of $\ln(1/x)$ for the same three inputs evaluated at 10, 100 and $10^4$~GeV, where the squared momentum-weighted correlation with $\sqrt{\ln(1/x)}$ is \NumSlopeSqrtXiHighMin{} to \NumSlopeSqrtXiHighMax{} and $(B/\kappa)^2$ is \NumSlopeXiHighMin{} to \NumSlopeXiHighMax{} (Sec.~\ref{sec:bound}).
The two descriptions agree because the momentum-weighted correlation between $\sqrt{\ln(1/x)}$ and $\ln(1/x)$ is strong for these densities, but large $x$ also contributes to this comparison, which does not test the small-$x$ form.
If $F_2$ is a single power of $1/x$ on a fixed window of $x$, as in the H1 fits~\cite{H1rise2001}, the speed of $F_2$ normalized on that window equals its mean bound for every value of the power.
This is the equality case of the bound, a property of the single-power form, and not a result about the H1 data.

The geometry also gives evolution a length and a distance between its end densities.
The lengths of the paths starting at the input scale and ending at $10^4$~GeV exceed the shortest connecting lengths by $\NumGeoExcessMin$ to $\NumGeoExcessMax$ percent for the five model inputs of Appendix~\ref{app:methods}, the three above and two with a hard quark or a hard gluon spectrum.
Near equality is a property of the rate of change at each scale and does not imply that the evolved densities of the proton-like, steep and valence-like inputs are reweightings of their inputs.
The proton-like density at $10^4$~GeV lies at Fisher-Rao distance $\NumGeoTiltDistA$ from the reweighting family through its input density.
Kou and Chen~\cite{KouChen2026} independently calculated the path length, the distance between its end densities, the excess of length over distance and the geodesic curvature for a different density, a normalized valence-quark number density evolved with the LO non-singlet kernel.

The quark and gluon input densities stay separately nonnegative through LO evolution, so $B$ bounds $\kappa$ at each scale of an LO path provided the other mean-bound assumptions are satisfied.
The two other model inputs give lower squared ratios, $\NumCorrSqExtraMinPct$ to $\NumCorrSqExtraMaxPct$ percent at the four scales, with the smallest for soft quarks and hard gluons at 2~GeV.
A scan of $\NumScanN$ inputs with one beta-distribution term per species gives $(B/\kappa)^2$ from $\NumScanMinTwo$ to $\NumScanMaxTwo$ at 2~GeV and from $\NumScanMinTenTeV$ to $\NumScanMaxTenTeV$ at $10^4$~GeV, reaching as low as $\NumScanMinQZero$ at the input scale.
The scan gives no lower limit that holds for all inputs (Sec.~\ref{sec:bound}~A), derivative exchange is checked only on the evaluated inputs (Table~\ref{tab:assumptions}), and the next-to-leading-order extension concerns only the drift identity (Appendix~\ref{app:nlo}).
Nonlinear small-$x$ terms are omitted, including the negative term quadratic in the gluon density that the Balitsky-Kovchegov equation~\cite{Balitsky1996,Kovchegov1999} gives for a large nucleus at large $Q^2$ (Eq.~(19) of Ref.~\cite{Kovchegov1999}).

We propose extending the width and mean bound using the known two-loop splitting functions~\cite{CurciFurmanskiPetronzio1980,FurmanskiPetronzio1980,Vogt2005} and the three-loop splitting functions of Moch, Vermaseren and Vogt~\cite{MochVermaserenVogt2004} and of Vogt, Moch and Vermaseren~\cite{VogtMochVermaseren2004}, and comparing the bound with speeds from fitted densities~\cite{CT18,NNPDF40,MSHT20} where positivity and derivative exchange hold.
A proposed test is to evaluate the excess of the normalized $F_2$ speed over its mean bound on a fixed window at the Large Hadron-Electron Collider described by Agostini et~al.~\cite{LHeC2021} or the Electron-Ion Collider described by Abdul Khalek et~al.~\cite{EICYR2022}. The excess would measure how closely the logarithmic rate of change of normalized $F_2$ follows a slope affine in $\ln(1/x)$, in mean square weighted by that normalized structure function.
A formula predicting the residual slope from the input and the LO kernel remains an open problem.

The main result of this paper is a lower bound in closed form for the momentum-density speed per unit of evolution time throughout leading-order DGLAP evolution with the flavor number fixed at five, splitting functions and running coupling at one loop, separate nonnegativity of the quark and gluon inputs, the LO moment assumptions together with the mean-bound assumptions, and positive logarithmic variance at every scale. The LO rates of change of the logarithms of the evolved momentum densities are close to slopes affine in $\ln(1/x)$ in mean square with momentum weighting for the proton-like, steep and valence-like inputs at 2, 10, 100 and $10^4$~GeV (and at the three higher scales closer still to slopes affine in $\sqrt{\ln(1/x)}$).
The three input numbers and the specified coupling thus give a bound within a few percent of the evolved-density speed for those three inputs at those four scales, with a square at least $\NumCorrSqMinPct$ percent of the speed squared.

\begin{acknowledgments}
The Anthropic Claude models listed in Appendix~\ref{app:ai} assembled and edited the text of the article and of the Supplemental Material.
Of the OpenAI models, GPT-6 Astra and GPT-6.1 Sol drafted and revised the prose, GPT-6 Astra also checked the text, and GPT-6 Luna retrieved literature.
Gemini 3.1 Pro (Google) and Kimi K3 (Moonshot AI) gave referee-style critique of the manuscript.
The use of AI tools in the research is described in Appendix~\ref{app:ai}.
\end{acknowledgments}

\appendix
\section{LO matrix and constants}
\label{app:matrix}

Momentum conservation makes the column sums of the singlet matrix vanish at $N=2$. Its derivatives at that point determine how the momentum distribution moves and widens in $\xi$. The space-like LO matrix in the coupling convention of Eq.~\eqref{eq:kappadefn} is~\cite{AltarelliParisi1977,Dokshitzer1977}
\begin{widetext}
\begin{equation}
\begin{aligned}
\Gamma_{qq}(N)&=C_F\left[\frac32+\frac{1}{N(N+1)}-2S_1(N)\right],\\
\Gamma_{qg}(N)&=2n_fT_R\,\frac{N^2+N+2}{N(N+1)(N+2)},\\
\Gamma_{gq}(N)&=C_F\,\frac{N^2+N+2}{(N-1)N(N+1)},\\
\Gamma_{gg}(N)&=2C_A\left[\frac{1}{N(N-1)}+\frac{1}{(N+1)(N+2)}-S_1(N)\right]\\
&\quad+\frac{11C_A-4n_fT_R}{6}.
\end{aligned}
\label{eq:gammaLO}
\end{equation}
\end{widetext}
Here $C_F=4/3$, $C_A=3$, $T_R=1/2$, and $S_1$ is the harmonic sum continued to complex Mellin argument. The column sums vanish at the momentum sum-rule point,
\begin{equation}
\begin{gathered}
c_q(2)=\Gamma_{qq}(2)+\Gamma_{gq}(2)=0,\\
c_g(2)=\Gamma_{qg}(2)+\Gamma_{gg}(2)=0 .
\end{gathered}
\label{eq:colsum}
\end{equation}
At this point the matrix and its eigenvalues are
\begin{equation}
\begin{gathered}
\Gamma(2)=\begin{pmatrix}-\frac{16}{9}&\frac{n_f}{3}\\[2pt]\frac{16}{9}&-\frac{n_f}{3}\end{pmatrix},\qquad
\gamma_+^{(0)}(2)=0,\\
\gamma_-^{(0)}(2)=-\frac{16}{9}-\frac{n_f}{3}=-\frac{31}{9}\quad(n_f=5).
\end{gathered}
\label{eq:gammaNtwo}
\end{equation}
Equation~\eqref{eq:cprime} gives the first column derivatives. The second derivatives are
\begin{equation}
c_q''(2)=\frac{16\,\zeta(3)}{3}-1,
\qquad
c_g''(2)=12\,\zeta(3)-\frac{395}{144}+\frac{71n_f}{432}.
\label{eq:csecond}
\end{equation}
The signs follow the splitting-function convention in Eq.~\eqref{eq:mellinevol}.

At $n_f=5$, the mean and variance rates per unit $\zeta$ are
\begin{equation}
\mu=\frac{408\pi^2-830}{2139}=\NumMu .
\label{eq:muclosed}
\end{equation}
\begin{equation}
D=\frac{1632\,\zeta(3)}{713}+\frac{8000\,\pi^4}{685193}+\frac{10000\,\pi^2}{685193}-\frac{2734750}{6166737}=\NumD .
\label{eq:Dclosed}
\end{equation}
The numerator of $\mu$ is positive. The next derivative gives $\Lambda_3=\NumLambdaThree$, while $D/\mu=\NumDoverMu$.

\begin{table*}[t]
\caption{Constants for the fixed-flavor LO evolution. The drift edges and the coefficient of the bound in $\tau$ do not assume one-loop running. The rates $\mu$, $D$, and $\Lambda_3$ of Eq.~\eqref{eq:muD} are per unit $\zeta$ of Eq.~\eqref{eq:clocks}, and $A_\kappa$ and $A_L$ are the coefficients in $\zeta$ of Eq.~\eqref{eq:bounds}. The drift edges are per unit $\tau$. The lower drift edge is \NumDriftLo{} for every flavor number shown.}
\label{tab:constants}
\begin{ruledtabular}
\begin{tabular}{lcccc}
Quantity & $n_f=3$ & $n_f=4$ & $n_f=5$ & $n_f=6$\\
\hline
$\mu$ per unit $\zeta$ & \NumMuNfThree & \NumMuNfFour & \NumMu & \NumMuNfSix\\
$D$ per unit $\zeta$ & \NumDNfThree & \NumDNfFour & \NumD & \NumDNfSix\\
$\Lambda_3$ per unit $\zeta$ & \NumLambdaThreeNfThree & \NumLambdaThreeNfFour & \NumLambdaThree & \NumLambdaThreeNfSix\\
$A_\kappa$ coefficient in $\zeta$ & \NumAkappaNfThree & \NumAkappaNfFour & \NumAkappa & \NumAkappaNfSix\\
$A_L$ coefficient in $\zeta$ & \NumALNfThree & \NumALNfFour & \NumAL & \NumALNfSix\\
$\lim\sqrt{\tau}\,B$ coefficient in $\tau$ & \NumTauConstNfThree & \NumTauConstNfFour & \NumTauConst & \NumTauConstNfSix\\
Upper drift edge per unit $\tau$ & \NumDriftHiNfThree & \NumDriftHiNfFour & \NumDriftHi & \NumDriftHiNfSix\\
\end{tabular}
\end{ruledtabular}
\end{table*}

Table~\ref{tab:constants} gives the flavor dependence. The figures use the fixed-flavor convention of Sec.~\ref{sec:density}. The last column gives the six flavors active above the top threshold.

\section{Cumulants at every scale}
\label{app:cumulants}

The width retains the separation of the quark and gluon means. The LO moment assumptions let us follow that separation without reconstructing the density in $x$. Define
\begin{equation}
\begin{gathered}
\Xi_{j,k}=\int_0^1 dx\,\xi^k P_j=\left(-\d_N\right)^k\tilde P_j(N)\Big|_{N=2},\\
\mathbf\Xi_k=\bigl(\Xi_{q,k},\Xi_{g,k}\bigr)^{\mathsf T},\qquad
G_k=\d_N^k\Gamma(N)\Big|_{N=2}.
\end{gathered}
\label{eq:momentdef}
\end{equation}
Differentiation of the matrix evolution gives a closed system at each finite moment order. Its first three equations are
\begin{equation}
\begin{gathered}
\d_\tau\mathbf\Xi_0=G_0\mathbf\Xi_0,\qquad
\d_\tau\mathbf\Xi_1=G_0\mathbf\Xi_1-G_1\mathbf\Xi_0,\\
\d_\tau\mathbf\Xi_2=G_0\mathbf\Xi_2-2G_1\mathbf\Xi_1+G_2\mathbf\Xi_0 .
\end{gathered}
\label{eq:momenteq}
\end{equation}
The species sum removes every term containing $G_0$, so the first-moment equation gives Eq.~\eqref{eq:drift}. Its integral with the momentum-fraction solution gives Eq.~\eqref{eq:meanxi}.

For the variance, the additional quantity is the quark contribution to the centered first moment,
\begin{widetext}
\begin{equation}
\begin{gathered}
C_q=\int_0^1 dx\,\bigl(\xi-\langle\xi\rangle\bigr)P_q=\langle x\rangle_q\langle x\rangle_g\bigl(\langle\xi\rangle_q-\langle\xi\rangle_g\bigr),\\
\langle\xi\rangle_j=\frac{\Xi_{j,1}}{\langle x\rangle_j}.
\end{gathered}
\label{eq:Cqdef}
\end{equation}
\end{widetext}
The integral defines $C_q$ even when one species fraction vanishes. The expression through separate species means applies when both fractions are nonzero. The variance obeys
\begin{widetext}
\begin{equation}
\frac{d\operatorname{Var}_P(\xi)}{d\tau}=c_q''(2)\,\langle x\rangle_q+c_g''(2)\,\langle x\rangle_g-2\bigl(c_q'(2)-c_g'(2)\bigr)\,C_q .
\label{eq:varrate}
\end{equation}
\end{widetext}
At $n_f=5$, $C_q$ obeys
\begin{equation}
\begin{gathered}
\frac{dC_q}{d\tau}=-\frac{31}{9}\,C_q+S,\\
S=\left(\frac{5\pi^2}{9}-\frac{17}{18}\right)\langle x\rangle_q^2-\left(\frac{5\pi^2}{9}+\frac{425}{216}\right)\langle x\rangle_q+\frac{55}{72}.
\end{gathered}
\label{eq:Cqrate}
\end{equation}

The coefficients in Eq.~\eqref{eq:varxi} are, at $n_f=5$,
\begin{equation}
\begin{aligned}
V_\infty&=\operatorname{Var}_0-\frac{10\pi^2-17}{31}\,C_{q0}-\eta^2+k_1\,X_0-\nu,\\
V_E&=\frac{10\pi^2-17}{31}\,C_{q0}+2\eta^2-k_1\,X_0+\nu,\\
V_{\zeta E}&=-\frac{5\bigl(24\pi^2+3859\bigr)}{12834}\,\eta,\\
k_1&=\frac{600\pi^4+95455\pi^2-691920\,\zeta(3)-68388}{357492},\\
\nu&=\frac{60\bigl(10\pi^2-17\bigr)\bigl(20\pi^2+59\bigr)}{31^4}=\frac{9}{31}\,F_{qg}\Big|_{n_f=5},\\
X_0&=\langle x\rangle_q(Q_0)-\frac{15}{31}.
\end{aligned}
\label{eq:varcoeffs}
\end{equation}
Here $\operatorname{Var}_0$ and $C_{q0}$ are the input variance and the input value of Eq.~\eqref{eq:Cqdef}. These two moments and the input quark momentum fraction fix the finite-scale variance, including the decaying eigenmode and its mixing with the growing cumulants. In $\nu$, $F_{qg}$ is the quark-gluon term of the limiting variance rate, defined in Eq.~\eqref{eq:memorysplit} of the Supplemental Material.

The drift interval in $\zeta$ units runs from \NumDriftLoZeta{} to \NumDriftHiZeta{} for one-loop running at $n_f=5$. High-precision differentiation of the generating function agrees with the moment solution for the tested inputs. A general multistate version of the finite-time generating-function method appears in Appendix~J and Eq.~(K1) of Ref.~\cite{Landi2024}. That work treats a broader quantum setting. The singlet coefficients above are calculated here.

\section{Derivation of the bound}
\label{app:bound}

The speed bound compares motion of the mean with the width of the distribution. It uses the mean-bound assumptions of Sec.~\ref{sec:density}, including $H_{\rm ex}$. Dominated convergence permits each derivative exchange if differentiability holds almost everywhere and an integrable, scale-independent majorant bounds the absolute derivative. The majorant must include the corresponding moment weight and hold in a neighborhood of the scale in question.

At an interior scale, the derivative of a nonnegative differentiable density vanishes almost everywhere on its zero set. Otherwise the density would become negative on one side of that scale. The normalization and first-moment identities can therefore be restricted to the positive set. Cauchy-Schwarz then gives
\begin{widetext}
\begin{equation}
\left|\frac{d\langle\xi\rangle}{d\tau}\right|
=\left|\int_{P>0}dx\,\bigl(\xi-\langle\xi\rangle\bigr)\sqrt P\,\frac{\d_\tau P}{\sqrt P}\right|
\le\operatorname{sd}_P(\xi)\,\sqrt{I_\tau}.
\label{eq:cauchyschwarz}
\end{equation}
\end{widetext}
Finite information is needed for this Hilbert-space argument. Positive information is also needed for the correlation in Eq.~\eqref{eq:identity}.

A boundary that moves with scale can invalidate the derivative exchange. The interior slope of a uniform density on a growing interval gives
\begin{equation}
\begin{gathered}
\rho(\xi,\tau)=\frac{1}{1+\tau}\quad(0<\xi<1+\tau),\\
\operatorname{Cov}(s_\tau,\xi)=0,\qquad
\frac{d\langle\xi\rangle}{d\tau}=\frac12,\qquad
\langle s_\tau\rangle=-\frac{1}{1+\tau}.
\end{gathered}
\label{eq:movingrange}
\end{equation}
The interior logarithmic slope misses the boundary contribution to the mean derivative. A moving endpoint does not by itself invalidate the identities. If the density vanishes sufficiently fast there, they may remain valid, as in
\begin{equation}
\begin{gathered}
P=\frac{3\bigl(x_{\max}-x\bigr)^2}{x_{\max}^3}\quad\bigl(0<x<x_{\max}(t)\bigr),\\
\langle s\rangle_P=0,\qquad
I_F=\frac{3\,\dot x_{\max}^2}{x_{\max}^2}.
\end{gathered}
\label{eq:vanishingend}
\end{equation}
Here $\dot x_{\max}=dx_{\max}/dt$. The relevant condition concerns the integral and its derivative. Specifying the region where the density is nonzero is insufficient.

Equality in Eq.~\eqref{eq:cauchyschwarz} at one scale requires a logarithmic slope affine in $\xi$ almost everywhere under the probability measure. Equality on an interval requires an exponential tilting family in $\xi$. General LO DGLAP evolution does not have this form. Integration gives
\begin{equation}
L(Q_0\to Q)\ge\int_{Q_0}^{Q}\frac{\bigl|d\langle\xi\rangle\bigr|}{\operatorname{sd}_P(\xi)},
\qquad
\int_0^\zeta\frac{\mu\,dz}{\sqrt{Dz}}=\frac{2\mu}{\sqrt D}\sqrt\zeta .
\label{eq:lengthbound}
\end{equation}
The first integral runs along the density path, with $Q_0$ and $Q$ labeling its end points. The second gives the leading large-time term. A fixed initial part of the path does not affect its coefficient, which yields the length limit in Eq.~\eqref{eq:bounds}.

The large-time bounds use finite moment equations, without a Gaussian density, saddle-point inversion, chosen eigenvector overlap, or further restriction on the input small-$x$ singularity. The coefficients depend on evolution time. In $\tau$, the explicit coupling is removed from each splitting, while conversion to $\zeta$ uses one-loop running. The speed and time increment change together in this conversion, so the length along the same path is unchanged. Its coefficient follows from the same drift and width, with no new coupling-dependent distance.

The approach of the speed to its lower bound is measured by
\begin{equation}
R=\frac{I_\zeta\operatorname{Var}_P(\xi)}{\bigl(d\langle\xi\rangle/d\zeta\bigr)^2}=\frac{1}{\operatorname{corr}_P(s,\xi)^2}\ge1 .
\label{eq:Rdef}
\end{equation}
This ratio is unchanged by the choice of evolution time or its reference scales along the same trajectory.

The variance change strengthens the mean bound through projection of the logarithmic slope onto the centered observables $\xi$ and $\xi^2$. Let $\mathbf m=(\langle\xi\rangle,\langle\xi^2\rangle)^{\mathsf T}$, let $\mathbf V$ be the covariance matrix of $\xi$ and $\xi^2$, and denote the cumulants by $\chi_n$. Primes denote $\zeta$ derivatives. If the quadratic-moment assumptions of Sec.~\ref{sec:density} hold, the moment-information inequality of Stein and Nossek~\cite{SteinNossek2017} and of Ito and Dechant~\cite{ItoDechant2020} gives
\begin{widetext}
\begin{equation}
\begin{gathered}
I_\zeta\ge J_2=\mathbf m'^{\mathsf T}\mathbf V^{-1}\mathbf m'
=\frac{\chi_1'^2}{\chi_2}+\frac{\bigl(\chi_2'-\chi_1'\chi_3/\chi_2\bigr)^2}{K_4},\\
K_4=\chi_4+2\chi_2^2-\frac{\chi_3^2}{\chi_2},
\end{gathered}
\label{eq:Jtwo}
\end{equation}
\begin{equation}
R\ge R_{\rm low}=\frac{J_2\,\chi_2}{\chi_1'^2}
=1+\frac{\chi_2\bigl(\chi_2'-\chi_1'\chi_3/\chi_2\bigr)^2}{\chi_1'^2K_4}.
\label{eq:Rlow}
\end{equation}
\end{widetext}
The first term in $J_2$ is the mean bound. The second measures the variance change left after accounting for the mean change and skewness.
The quadratic bound uses the moment-information inequality of Ref.~\cite{SteinNossek2017}, where Jarrett~\cite{Jarrett1984} is cited for an alternative using raw moments and cumulants.

Finite information also restricts the input density. At the input scale it fails when both quark and gluon momentum densities vanish at least as fast as $x$ at small $x$. At the upper endpoint, take a quark density proportional to $(1-x)^b$ and a gluon density with a strictly larger endpoint exponent. The known LO endpoint behavior given by M\"uller~\cite{Mueller1994} and by Abdulov, Kotikov and Lipatov~\cite{AKL2022} gives
\begin{equation}
\begin{gathered}
s_\tau(x)=2C_F\bigl[\ln(1-x)-\psi(b+1)\bigr]\\
+C_F\left(\frac32-2\gamma_E\right)+o(1),\qquad x\to1 .
\end{gathered}
\label{eq:xtoone}
\end{equation}
Here $\psi$ is the digamma function and $\gamma_E$ is Euler's constant. The logarithmic growth leaves the upper endpoint of the Fisher integral convergent for every integrable quark power. This use of Eq.~\eqref{eq:xtoone} requires the stated quark dominance.

\section{Markov form and thresholds}
\label{app:markov}

At LO, real emissions redistribute momentum and virtual terms supply the compensating loss. LO evolution is a master equation with explicit gain and loss terms, as in the work of Wosiek and Zalewski~\cite{WosiekZalewski1979}, Durand and Putikka~\cite{DurandPutikka1987}, and Collins and Qiu~\cite{CollinsQiu1989}, and in Ref.~\cite{Teryaev1998}, written for number densities in Refs.~\cite{WosiekZalewski1979,DurandPutikka1987} and for the momentum-weighted densities in Ref.~\cite{Teryaev1998}. Cafarella and Corian\'o~\cite{CafarellaCoriano2005} study the kinetic interpretation and its Kramers-Moyal expansion in $x$. Denote the species different from $a$ by $a'$. The Mellin entries are
\begin{widetext}
\begin{equation}
\begin{aligned}
\Gamma_{ab}(N)&=\int_0^1 dz\,zP_{ab}(z)\,z^{N-2}\qquad(a\neq b),\\
\Gamma_{aa}(N)&=\int_0^1 dz\,zP^{\rm R}_{aa}(z)\bigl(z^{N-2}-1\bigr)-\int_0^1 dz\,zP_{a'a}(z).
\end{aligned}
\label{eq:gainloss}
\end{equation}
\end{widetext}
The LO real-emission rates are nonnegative. The diagonal subtraction compensates the soft singularity, and the remaining loss accounts for transitions to the other species. Momentum conservation fixes the total column mass.

The logarithmic momentum loss $\ell$ and species label define a Markov additive process. Its matrix exponent is $-\Gamma(2+\theta)^{\mathsf T}$ in the multitype fragmentation convention of Ref.~\cite{Bertoin2008}. The transpose is needed because that convention~\cite{Bertoin2008} and Eq.~\eqref{eq:mellinevol} act on opposite sides of the species vector. The resulting DGLAP transition measures are nonnegative and have the transform in Eq.~\eqref{eq:markov}.

The logarithmic derivative at an output point averages the input logarithmic derivatives if the transition kernel is scale-independent. The weights are proportional to the input density times the transition probability to that point. Jensen's inequality contracts the squared mean and gives Fisher-information data processing~\cite{AJLS2015}. A common further evolution of neighboring points on a time-homogeneous path gives Eq.~\eqref{eq:jointmono} if the Markov-contraction assumptions hold. The species sum gives Eq.~\eqref{eq:jointgap}, without requiring a Markov marginal process.

A heavy-flavor threshold opens a new channel. If the LO matching assumptions of Sec.~\ref{sec:density} hold, the singlet matrix changes by
\begin{equation}
\begin{gathered}
\Delta\Gamma(N)=\begin{pmatrix}0&2\Gamma_{hg}(N)\\0&-\frac13\end{pmatrix},\qquad
2\Gamma_{hg}(2)=\frac13,\\
\Gamma_{hg}(N)=T_R\,\frac{N^2+N+2}{N(N+1)(N+2)}.
\end{gathered}
\label{eq:dGamma}
\end{equation}
The total kernel action changes by
\begin{equation}
\begin{gathered}
\Delta K=2\bigl[zP_{hg}\otimes xg\bigr]-\frac13\,xg,\qquad
\int_0^1\Delta K\,dx=0,\\
P_{hg}(z)=T_R\bigl[z^2+(1-z)^2\bigr].
\end{gathered}
\label{eq:deltaK}
\end{equation}
The momentum sum rule makes the integral vanish. Expansion of the squared slope gives Eq.~\eqref{eq:thresholdjump}. The first logarithmic moment of $\Delta K$ gives Eq.~\eqref{eq:thresholddrift}. The drift jump needs the finite total first logarithmic moment and derivative exchange specified in the LO matching assumptions. A finite gluon logarithmic moment alone is insufficient.

The information integral changes when the emerging heavy flavor is resolved separately. If its input density vanishes and the gluon momentum fraction at the threshold is nonzero,
\begin{equation}
I_{\rm flavor}(\tau)=\frac{\langle x\rangle_g}{3\,(\tau-\tau_{\rm th})}+O(1),
\qquad
\tau\to\tau_{\rm th}^+ .
\label{eq:heavydiv}
\end{equation}
The divergence comes from the emerging heavy density. It is absent when that density belongs to an already present quark singlet. Both singlet-gluon and marginal information remain finite if the species-resolved Fisher information at the input is finite, as required by the LO matching assumptions of Sec.~\ref{sec:density}. The main-text jump formulas concern the marginal density.

\section{Path geometry and the approach to the bound}
\label{app:path}

The amplitude $u=2\sqrt P$ lies on the sphere of radius 2 in the square-integrable functions with inner product $\langle f,g\rangle=\int_0^1dx\,fg$ and norm $\|f\|=\langle f,f\rangle^{1/2}$, along a path of normalized nonnegative densities on $0<x<1$ with measure $dx$,
\begin{equation}
u(x,t)=2\sqrt{P(x,t)},\qquad
\int_0^1 dx\,u^2=4,\qquad
\int_0^1 dx\,\bigl(\d_tu\bigr)^2=I_F(t).
\label{eq:sphere}
\end{equation}
The third relation needs the Fisher assumptions and $u(t)$ absolutely continuous in the square-integrable functions, with finite length. Its derivative is $\d_tu=\sqrt P\,s$ where $P>0$ and vanishes almost everywhere where $P=0$, as in Appendix~\ref{app:bound}. Thus the metric $ds^2=I_F\,dt^2$ of Sec.~\ref{sec:density}~B is the metric of this sphere, and differentiation of $\|u\|^2=4$ makes $\d_tu$ orthogonal to $u$. Define $\phi(t)$ by $\cos\phi(t)=\langle u(t_1),u(t)\rangle/4$. The component of $u(t_1)$ orthogonal to $u(t)$ has norm $2\sin\phi$, so Cauchy-Schwarz gives $|\d_t\phi|\le\|\d_tu\|/2$ wherever $\sin\phi>0$. Hence $L\ge2\phi(t_2)=D_{\rm FR}$. Equality requires velocity parallel to that component almost everywhere and nondecreasing $\phi$, giving the great-circle arc without reversal,
\begin{equation}
\sqrt{P_\sigma}=\frac{\sin[(1-\sigma)\phi]\sqrt{P(t_1)}+\sin(\sigma\phi)\sqrt{P(t_2)}}{\sin\phi},\qquad
\cos\phi=\int_0^1dx\,\sqrt{P(t_1)\,P(t_2)},\qquad
0\le\sigma\le1 .
\label{eq:geodesic}
\end{equation}
Here $\phi=\phi(t_2)$. The overlap $\cos\phi$ is nonnegative, so $D_{\rm FR}\le\pi$ and the end points are never antipodal. For coincident densities, the arc is defined by its constant-path limit.

The acceleration is $\d_t^2u=\sqrt P\,(\d_ts+s^2/2)$ in any path parameter $t$. Equation~\eqref{eq:secondnorm} gives its radial component $-I_Fu/4$ and tangential part $\sqrt P\,w$. Removing from $\sqrt P\,w$ its projection on the velocity $\sqrt P\,s$ and dividing the squared norm by the fourth power of the speed gives the squared geodesic curvature, which does not depend on the parameter,
\begin{equation}
\begin{gathered}
k^2=\frac{1}{I_F^2}\left[\bigl\langle w^2\bigr\rangle_P-\frac{\langle ws\rangle_P^2}{I_F}\right],\qquad
w=\frac{\d_t^2P}{P}-\frac{s^2}{2}+\frac{I_F}{2},\\
\langle w\rangle_P=0,\qquad
\langle ws\rangle_P=\frac12\,\frac{dI_F}{dt}.
\end{gathered}
\label{eq:curvature}
\end{equation}
These relations assume $P>0$ almost everywhere on $0<x<1$, $u(t)$ twice continuously differentiable in the square-integrable functions, the derivative exchanges of Eq.~\eqref{eq:secondnorm}, positive $I_F$, and finite $\langle w^2\rangle_P$. For Eq.~\eqref{eq:curvlaw}, assume four $t$ derivatives of $u$ in this space. Expansion of $\langle u(-h),u(h)\rangle$ about the midpoint gives Eq.~\eqref{eq:curvlaw} if arc length is the parameter. The sphere radius cancels. The remainder is $O(L^4)$ with $k$ at the midpoint and $O(L^3)$ with $k$ at an endpoint.

The reweighting path through the first density is
\begin{equation}
P_\theta(x)=\frac{x^{-\theta}P(x,t_1)}{M_1(\theta)},\qquad
M_1(\theta)=\int_0^1dx\,x^{-\theta}P(x,t_1),
\label{eq:tiltpath}
\end{equation}
taken for $\theta\ge0$ where $M_1$ is finite. At $Q_0$, this requires $\theta<1+a_{\min}$ if the smallest input power is $x^{a_{\min}}$. This is the exponential tilting family in $\xi$ of Appendix~\ref{app:bound}. The slope per unit $\theta$ is $\xi-d\ln M_1/d\theta$, so its information is the variance of $\xi$ under $P_\theta$, $d^2\ln M_1/d\theta^2$. Its length, distance from the starting density, and curvature are
\begin{equation}
\begin{gathered}
L_{\rm tilt}(\theta)=\int_0^\theta d\theta'\,\sqrt{\frac{d^2\ln M_1}{d\theta'^2}},\qquad
\cos\frac{D_{\rm FR}}{2}=\frac{M_1(\theta/2)}{\sqrt{M_1(\theta)}},\\
k_{\rm tilt}^2=\frac{K_4}{4\chi_2^2}.
\end{gathered}
\label{eq:tiltlength}
\end{equation}
Here $\chi_2$ and $K_4$ of Eq.~\eqref{eq:Jtwo} are evaluated with $P_\theta$. Cauchy-Schwarz gives $K_4\ge0$, and a Gaussian in $\xi$ gives $k_{\rm tilt}^2=1/2$.

The weight $\mathcal T$ in Eq.~\eqref{eq:betaweight} satisfies $\mathcal T'=-(\xi-\langle\xi\rangle)p$, with vanishing endpoint values and one change of derivative sign, so $\mathcal T\ge0$. Integration by parts in $\operatorname{Cov}_P(s_\tau,\xi)$ with $\d_\xi s_\tau=\d_\tau\lambda$ gives the weighted integral, equal to $d\langle\xi\rangle/d\tau=\beta\operatorname{Var}_P(\xi)$ by Eq.~\eqref{eq:covdrift}. The same integration with $\xi-\langle\xi\rangle$ gives $\int\mathcal T\,d\xi=\operatorname{Var}_P(\xi)$. This derivation assumes $P>0$ on $0<x<1$, $\ln P$ twice continuously differentiable in $\xi$ and $\tau$, $s_\tau\mathcal T$ vanishing at both ends, and finite $\int|\d_\tau\lambda|\mathcal T\,d\xi$.

If both input species vanish like positive powers of $x$ at small $x$, as beta-distribution inputs with positive powers do, their Mellin transforms are regular at $N=1$. Let $P(x,0)\sim Ax^a$ as $x\to0$, with $0<a<1$ the smallest power of $x$ in the two species and $A$ the sum of their coefficients of that power. The transform $(1,1)\Gamma(N)\tilde{\mathbf P}(N,0)$ of $\d_\tau P$ has a simple pole there from $\Gamma_{gq}$ and $\Gamma_{gg}$ in the gluon row of Eq.~\eqref{eq:gammaLO}, whose residues are $2C_F$ and $2C_A$. Its residue is $\mathcal R$ of Eq.~\eqref{eq:transient}, so $\d_\tau P(x,0)\to\mathcal R$ as $x\to0$ and $s_\tau\sim(\mathcal R/A)x^{-a}$. For small positive $\tau$, $P\sim Ax^a+\tau\mathcal R$ cuts off this large slope at $x_c=(\tau\mathcal R/A)^{1/a}$, requiring $x_c\ll1$ and $\tau\xi\ll1$ there. The valence-like input has the power $a=1/2$, for which
\begin{equation}
\begin{gathered}
I_\tau(\tau)-I_\tau(0)=-\frac{2\mathcal R^3}{A^2}\,\tau\ln\frac1\tau+O(\tau),\\
\mathcal R=2\bigl[C_A\tilde P_g(1,0)+C_F\tilde P_q(1,0)\bigr],\qquad
P(x,0)=A\,x^{1/2}\bigl[1+O(x^d)\bigr],\ d>0,\ x\to0 .
\end{gathered}
\label{eq:transient}
\end{equation}
This result follows by inserting the small-$x$ form into $I_\tau$ and subtracting $I_\tau(0)$. The input transforms must be regular to the left of $N=1$, as for the beta inputs. In contrast, $B^2$ has a power series in $\tau$, since its moments are $N$ derivatives of $(1,1)e^{\tau\Gamma(N)}\tilde{\mathbf P}(N,0)$ at $N=2$. The deficit's leading change is therefore $(B^2/I_\tau^2)[I_\tau(\tau)-I_\tau(0)]$. For $1/2<a<1$ and $d>2a-1$, the corresponding law is $I_\tau(\tau)-I_\tau(0)=-c_a\mathcal R^{1+1/a}A^{-1/a}\tau^{(1-a)/a}+O(\tau)$, where $c_a=\pi/[a\sin(\pi(1-a)/a)]$.

The valence-like input has $a=1/2$ from the gluon. The quark term, of power $x^{\NumInValenceQa}$, gives the correction with $d$ the difference of the quark and gluon powers. The constants of the valence-like input are $\mathcal R=\NumTransR$, $A=\NumTransA$, and coefficient magnitude $2\mathcal R^3/A^2=\NumTransCoef$. Equation~\eqref{eq:transient} applies only at the beginning of the transient. The later approach is described only by the computed valence-like deficit in the lower panel of Fig.~\ref{fig:bound} and in Sec.~\ref{sec:bound}~A.

\section{NLO drift at \texorpdfstring{$N=2$}{N=2}}
\label{app:nlo}

Momentum conservation still makes the column sums vanish at NLO. Their derivatives therefore give the drift with the two-loop correction included. The Mellin method has been used for time-like cumulants~\cite{FongWebber1991,LupiaOchs1996,KhozeOchs1997,PerezRamosdEnterria2014} and for the space-like non-singlet first moment~\cite{Artru2009}. Here we use the NLO moment assumptions collected in Sec.~\ref{sec:density}. The coupled momentum density obeys
\begin{equation}
\begin{gathered}
\frac{d\langle\xi\rangle}{dt}=-a_s\sum_{j=q,g}\Bigl[c_{0,j}'+a_s\,c_{1,j}'\Bigr]\langle x\rangle_j,\\
c_{n,j}'=\sum_i\d_N\Gamma_{ij}^{(n)}(N)\Big|_{N=2},\qquad
a_s=\frac{\alpha_s}{2\pi}.
\end{gathered}
\label{eq:nlodrift}
\end{equation}
The two-loop matrix follows the \mbox{PEGASUS} transcription~\cite{Vogt2005} of the two-loop splitting functions~\cite{CurciFurmanskiPetronzio1980,FurmanskiPetronzio1980}. The \mbox{PEGASUS} transcription~\cite{Vogt2005} expands in $\alpha_s/(4\pi)$, so $\Gamma^{(1)}$ here is one quarter of its two-loop matrix.

\begin{table}[t]
\caption{Two-loop column derivatives $c_{1,q}'$ and $c_{1,g}'$ of Eq.~\eqref{eq:nlodrift} for $n_f=4$ and 5, with $a_s=\alpha_s/(2\pi)$.}
\label{tab:nlocolumns}
\begin{ruledtabular}
\begin{tabular}{ccc}
$n_f$ & $c_{1,q}'$ & $c_{1,g}'$\\
\hline
$4$ & \NumCOnePrimeSigmaNfFour & \NumCOnePrimeGNfFour\\
$5$ & \NumCOnePrimeSigma & \NumCOnePrimeG\\
\end{tabular}
\end{ruledtabular}
\end{table}

Table~\ref{tab:nlocolumns} gives the column derivatives, and Table~\ref{tab:nlotest} gives the drift comparison for the central density of the NLO fit of CT18~\cite{CT18}. Each residual $r_{\rm NLO}$ is the grid drift divided by the identity's drift, minus one. The comparison tests the implementation, including column derivatives, coupling conventions, and the grid calculation, without testing QCD dynamics independently. The grid drift is evaluated on $[10^{-9},0.99]$, while the identity applies to the full interval. For the grid drift, we differentiate a bicubic spline of the densities in $\ln x$ and $\ln Q$ with respect to $t$. The spline uses only the tabulated scales of the grid in the same fixed-flavor segment as $Q$, since the identity holds within such a segment. The fitted speeds instead use the spline through every grid node described in Sec.~\ref{sm:fitmethods} of the Supplemental Material. The momentum fractions in the identity are integrated over $[10^{-9},1-10^{-7}]$ and are not rescaled to make their sum one. For the check, $\alpha_s$ is taken from the $\alpha_s$ table of the same CT18 set~\cite{CT18} and interpolated in $\ln Q$. The largest residual among \NumAOneScanN{} scales from 2 to 100~GeV is \NumRNLOScan{} percent at \NumRNLOScanQ{}~GeV, above the CT18 bottom threshold~\cite{CT18} $m_b=\NumMbCT$~GeV.

\begin{table*}[t]
\caption{Drift check on the central member of the NLO fit of CT18~\cite{CT18}. The grid drift is evaluated on $[10^{-9},0.99]$, and the momentum fractions in the identity are integrated over $[10^{-9},1-10^{-7}]$. The NLO residual is bounded by \NumRNLOMax{} percent at the displayed scales. The scan contains \NumAOneScanN{} scales.}
\label{tab:nlotest}
\begin{ruledtabular}
\begin{tabular}{ccc}
$Q\;(\mathrm{GeV})$ & NLO residual (percent) & Residual with only the LO term (percent)\\
\hline
$2$ & \NumRNLOTwo & \NumRLOTwo\\
$5$ & \NumRNLOFive & \NumRLOFive\\
$10$ & \NumRNLOTen & \NumRLOTen\\
$100$ & \NumRNLOHundred & \NumRLOHundred\\
\end{tabular}
\end{ruledtabular}
\end{table*}

Let $v_+$ be the null right eigenvector of $a_s\Gamma^{(0)}(2)+a_s^2\Gamma^{(1)}(2)$ at the current coupling, normalized to $\langle x\rangle_q+\langle x\rangle_g$. The asymptotic part is Eq.~\eqref{eq:nlodrift} with $v_+$ in place of the momentum vector and equals the larger-eigenvalue derivative term. The finite-scale part is the remainder from the decaying eigenvector. It is negative, with magnitude \NumFinitePartTwo{} percent of the full NLO prediction at 2~GeV and \NumFinitePartHundred{} percent at 100~GeV.

At 10 and 100~GeV the NNLO grid of CT18~\cite{CT18} also satisfies the NLO identity to \NumAOneNNLOGridMax{} percent, so the check resolves the two-loop term but not the three-loop term there.

\section{Evaluation methods and reproducibility}
\label{app:methods}

The moment equations give the drift and width without reconstructing the density. Mellin inversion gives the density and its Fisher speed for the same evolution. Fitted grids need separate prescriptions for derivatives and uncertainties, and those for the fitted parton densities and for the structure-function data are given in the Supplemental Material.

\subsection{Inputs and evolution}

The model inputs let us vary the momentum at small $x$ while preserving the momentum sum. The proton-like input is
\begin{widetext}
\begin{equation}
\begin{aligned}
x\Sigma={}&0.38\,\frac{x^{0.5}(1-x)^3}{\mathrm{Beta}(1.5,4)}\\
&+0.20\,\frac{x^{-0.15}(1-x)^7}{\mathrm{Beta}(0.85,8)},\\
xg={}&0.42\,\frac{x^{-0.1}(1-x)^5}{\mathrm{Beta}(0.9,6)} .
\end{aligned}
\label{eq:inputs}
\end{equation}
\end{widetext}
Here $\mathrm{Beta}$ is the beta function. Each term is $c\,\mathcal B_{a,b}(x)$, where $\mathcal B_{a,b}(x)=x^a(1-x)^b/\mathrm{Beta}(a+1,b+1)$, so its momentum fraction is $c$. Table~\ref{tab:settings} gives the starting scale, coupling, and fixed flavor number.

For the model inputs, $\Lambda_{\rm QCD}$ enters only the map from $Q$ to $\tau$ and $\zeta$. The alternative value $\Lambda_{\rm QCD}=\NumLambdaAlt$~GeV in the one-loop coupling gives $\alpha_s(M_Z)=\NumAlphasMZAlt$, lowers $\zeta$ at 10, 100, and $10^4$~GeV from \NumZetaTen, \NumZetaHundred, and \NumZetaTenTeV{} to \NumZetaAltTen, \NumZetaAltHundred, and \NumZetaAltTenTeV{}, and moves the model curves to higher $Q$. The fixed-flavor LO curves and speeds per unit $\tau$ or $\zeta$ are unchanged as functions of those times. Speeds per unit $\ln Q^2$ retain the coupling factor and change.

The steep input has $x\Sigma=\NumInSteepQOnec\,\mathcal B_{\NumInSteepQOnea,3}+\NumInSteepQTwoc\,\mathcal B_{\NumInSteepQTwoa,7}$ and $xg=\NumInSteepGc\,\mathcal B_{\NumInSteepGa,5}$. The valence-like input, which has no small-$x$ rise, has $x\Sigma=\NumInValenceQc\,\mathcal B_{\NumInValenceQa,3}$ and $xg=\NumInValenceGc\,\mathcal B_{\NumInValenceGa,5}$. Their quark momentum fractions are listed in Table~\ref{tab:settings}.

We invert the Mellin solution by the Talbot method with the precisions and term counts in Table~\ref{tab:settings}. At low scale, the higher-precision valence-like calculation reproduces the lower-precision printed digits and has no negative species value at an integration node. The proton-like input density is also evaluated at the higher precision. The finite moment formulas are evaluated independently.

\begin{table*}[t]
\caption{Definitions and computational settings. The model LO trajectories use fixed $n_f=5$ throughout their scale range, as specified in Sec.~\ref{sec:density}. The LO trajectory from CT18~\cite{CT18} has the separate coupling and threshold settings below. The fitted-band confidence level also applies to Table~\ref{tab:fitted}.}
\label{tab:settings}
\begin{ruledtabular}
\begin{tabular}{ll}
Setting & Value\\
\hline
QCD color factors & $C_F=4/3,\ C_A=3,\ T_R=1/2$\\
LO starting scale and coupling parameter & $Q_0^2=2\ \mathrm{GeV}^2,\quad \Lambda_{\rm QCD}=\NumLambdaQCD\ \mathrm{GeV}$\\
LO flavor number & $n_f=5$\\
Steep and valence-like input quark fractions & \NumQuarkFracSteep,\quad\NumQuarkFracValence\\
Talbot working digits and terms & \NumTalbotDigits,\ \NumTalbotTerms\\
Higher-precision working digits and terms & \NumTalbotDigitsHigh,\ \NumTalbotTermsHigh\\
Figure band central value & \begin{tabular}[t]{@{}l@{}}Scan scales in each interval, with the Table~\ref{tab:fitted} values\\at 2 and 10~GeV\end{tabular}\\
Figure band relative uncertainty & Linear interpolation in $\ln Q$ between the uncertainty scales\\
Figure band uncertainty scales $Q\;(\mathrm{GeV})$ & $2,\ 10,\ 100,\ 10^3,\ 10^4$\\
Figure band relative uncertainty at these scales (percent) & \NumBandUncTwo,\ \NumBandUncTen,\ \NumBandUncHundred,\ \NumBandUncThousand,\ \NumBandUncTenTeV\\
Figure band intervals $Q\;(\mathrm{GeV})$ & $[2,\NumBandFirstEnd],\quad[10,10^4]$\\
CT18~\cite{CT18} bottom threshold & $m_b=\NumMbCT\ \mathrm{GeV}$\\
Fitted-band confidence level & $68\ \mathrm{percent}$\\
CT18~\cite{CT18} Hessian pairs and input confidence level & \NumCTHessianPairs,\quad \NumCTHessianCL{} percent\\
Coupling-variation sets & $\alpha_s(M_Z)=\NumAlphasLow,\quad\NumAlphasHigh$\\
CT18~\cite{CT18} threshold scale scan from \NumThresholdScanLo{} to $10^4$~GeV & \NumThresholdScanN{} points, geometric in $Q$\\
Small-$x$ share comparison & $x<10^{-2},\quad Q=10\ \mathrm{GeV}$\\
Extended-window lower endpoint & $x_{\min}=10^{-9}$\\
Low-scale interpolation comparison & $Q=\NumInterpLowQ\ \mathrm{GeV}$\\
LO trajectory from CT18~\cite{CT18}, starting scale & $Q=2\ \mathrm{GeV}$\\
LO trajectory from CT18~\cite{CT18}, evolution time & $\tau$ from the $\alpha_s$ table of that fit\\
LO trajectory from CT18~\cite{CT18}, flavor thresholds & $n_f=4$ below $m_b=\NumMbCT\ \mathrm{GeV}$ and 5 above, no top threshold\\
Additional-input squared correlations & \begin{tabular}[t]{@{}l@{}}Hard-quark, steep-gluon input\\\NumCorrSqHardqSteepgMin{} to \NumCorrSqHardqSteepgMax{} at $Q=2,\ 10,\ 100,\ 10^4\ \mathrm{GeV}$\\smallest, \NumCorrSqHardqSteepg{}, at $Q=10\ \mathrm{GeV}$ (Fig.~\ref{fig:bound})\\Soft-quark, hard-gluon input\\\NumCorrSqSoftqHardgQZero{} at $Q_0$\\\NumCorrSqSoftqHardg,\ \NumCorrSqSoftqHardgTen,\ \NumCorrSqSoftqHardgHundred,\ \NumCorrSqSoftqHardgTenTeV{}\\at $Q=2,\ 10,\ 100,\ 10^4\ \mathrm{GeV}$, respectively\end{tabular}\\
Window comparison with the quadratic bound & $Q=100,\ 10^4\ \mathrm{GeV}$\\
\end{tabular}
\end{ruledtabular}
\end{table*}

The additional nonnegative inputs test how far the correlation can fall below unity. The hard-quark, steep-gluon input has $x\Sigma=\NumInHqSgQc\,\mathcal B_{\NumInHqSgQa,3}$ and $xg=\NumInHqSgGc\,\mathcal B_{\NumInHqSgGa,8}$. The soft-quark, hard-gluon input has $x\Sigma=\NumInSqHgQc\,\mathcal B_{\NumInSqHgQa,7}$ and $xg=\NumInSqHgGc\,\mathcal B_{\NumInSqHgGa,4}$. Table~\ref{tab:settings} gives the complete evaluated squared correlations. They rise monotonically from 2 to $10^4$~GeV for the soft-quark, hard-gluon input.

The LO trajectory from CT18~\cite{CT18} starts from its central NNLO density at $Q=2$~GeV. We retain its tabulation on $10^{-6}\le x\le\NumCTContXu$, continue it below $10^{-6}$ by its local power of $x$, and above \NumCTContXu{} by $(1-x)^b$ with $b$ matched to the local logarithmic derivative. Division by the momentum sum normalizes the result. Moving the lower continuation point to $10^{-5}$ or $10^{-7}$, or changing the continued power by $\pm\NumCTContPowShift$, changes $\kappa$ at $Q=2$, 5, 10, 100, and $10^4$~GeV. The largest relative change, rounded to two significant figures, is \NumCTContSens{} and occurs at 2~GeV. We evaluate $B$ on this trajectory from Eq.~\eqref{eq:identity} using the moments of the evolved density and their scale derivative, since the closed form of Eq.~\eqref{eq:Bzeta} holds at fixed $n_f=5$ and the trajectory crosses the bottom-quark threshold.

\subsection{Reproduction material}

Reproduction requires the density, its derivative prescription, and the uncertainty calculation. The programs that compute every number of this article and the values plotted in its figures, with their stored outputs, are available as ancillary files of the arXiv version of this article~\cite{AncillaryFiles}. These files include the evolution and quadrature programs, fitted-grid versions and checksums, pseudo-data seed, and stored belt and coverage results.
The ancillary files also contain the programs and stored results for the input scan and the residual shares of Sec.~\ref{sec:bound}, the overlaps, path lengths and curvatures of Appendix~\ref{app:path}, and the half-deficit scale of the valence-like input.
They also include the additional input profiles and continuation of CT18~\cite{CT18}. The parton-distribution grids are available through the cited fit publications and LHAPDF~\cite{LHAPDF}, and the HERA measurements with their correlated errors through the combined H1 and ZEUS release~\cite{HERAcomb}. These third-party data are not redistributed in the ancillary files.

\begin{sloppypar}
The LHAPDF sets are \texttt{CT18NNLO}, \texttt{CT18NNLO\_as\_0116}, \texttt{CT18NNLO\_as\_0120}, \texttt{CT18NLO} and \texttt{CT18LO} for CT18~\cite{CT18}, \texttt{MSHT20nnlo\_as118} for MSHT20~\cite{MSHT20}, \texttt{NNPDF40\_nnlo\_as\_01180} for NNPDF4.0~\cite{NNPDF40}, and \texttt{NNPDF31sx\_nnlo\_as\_0118} and \texttt{NNPDF31sx\_nnlonllx\_as\_0118} for the NNPDF3.1sx pair~\cite{BallResum2018}. Member 0 is the central member. The programs generate \texttt{numbers.tex}, through which computed values enter the text. 
\end{sloppypar}

\subsection{Use of AI tools}
\label{app:ai}

We used the Anthropic models Claude Opus 4.8, Claude Opus 5, Claude Opus 5.5, Claude Fable 5, Claude Sonnet 5, Claude Sonnet 5.5, and Claude Haiku 4.5.
These models developed results and derivations with the authors, wrote and ran the programs (including those drawing the figures), and ran checks.
Of the OpenAI models, GPT-6 Astra independently checked derivations and numbers, and GPT-5.6 Sol made independent checks.
The uses for the text and the literature are given in the Acknowledgments.
The Claude and GPT models worked on the authors' instructions.
Gemini 3.1 Pro (Google) and Kimi K3 (Moonshot AI) were used only for checks and for the critique stated in the Acknowledgments.
The checks included computer algebra, evaluation at sample values, and independent recomputation of selected results.

Figures~\ref{fig:drift} and~\ref{fig:bound} of this paper and Fig.~\ref{fig:hera} of the Supplemental Material are deterministic ROOT plots drawn by these programs from stored computed results.
No image was generated or modified by a generative model.
The authors checked the plotting programs and the plotted values against the computed results.
The authors reviewed the text, derivations, programs, and figures, confirm that the manuscript reflects their work, and take full responsibility for its content.
No AI tool is an author.

\makeatletter
\let\auto@bib@innerbib\@empty
\makeatother
\microtypesetup{protrusion=false}

\end{document}